\documentclass[aps,a4paper,floatfix,nofootinbib,unsortedaddress]{revtex4}

\usepackage{graphicx,color}
\usepackage{amsmath,amssymb,amsfonts,amsthm,amscd,bm}
\usepackage{booktabs}
\usepackage[english]{babel}
\usepackage{mathtools}
\usepackage{mathrsfs}
\usepackage{bbm}
\usepackage{physics}
\usepackage{enumerate}
\usepackage{multirow}
\usepackage{xcolor}
\usepackage{listings}
\usepackage{pgfplots}
\usepackage[makeroom]{cancel}
\usepackage{tikz}
\usepackage[compat=1.1.0]{tikz-feynman}
\usetikzlibrary{positioning}
\usetikzlibrary{decorations.pathreplacing}
\usetikzlibrary{calc,angles,positioning,intersections,quotes,decorations.markings}
\usetikzlibrary{babel}
\usepackage{subfig}
\usepackage{array}
\graphicspath{{../pictures/}}
\newcommand{\comment}[1]{}
\comment{
\def\tilde{\widetilde}
\def\bar{\overline}
\def\hat{\widehat}
\def\*{\star}
\def\[{\left[}
\def\]{\right]}
\def\({\left(}      

\def\){\right)}

\def\frac#1#2{\dfrac{#1}{#2}}

\def\dd#1#2{{\partial #1 \over \partial #2}}

\def\2pi{\hbox{$2\pi i$}}

\def\dsl{\raise.15ex\hbox{/}\kern-.57em\partial}
\def\Dsl{\,\raise.15ex\hbox{/}\mkern-.13.5mu D}
\def\2pi{\hbox{$2\pi i$}}

\def\dsl{\raise.15ex\hbox{/}\kern-.57em\partial}
\def\Dsl{\,\raise.15ex\hbox{/}\mkern-.13.5mu D}
\font\numbers=cmss12
\font\upright=cmu10 scaled\magstep1
\def\stroke{\vrule height8pt width0.4pt depth-0.1pt}
\def\topfleck{\vrule height8pt width0.5pt depth-5.9pt}
\def\botfleck{\vrule height2pt width0.5pt depth0.1pt}
\def\Zmath{\vcenter{\hbox{\numbers\rlap{\rlap{Z}\kern
    0.8pt\topfleck}\kern 2.2pt
    \rlap Z\kern 6pt\botfleck\kern 1pt}}}
\def\Qmath{
    \vcenter{\hbox{\upright\rlap{\rlap{Q}\kern3.8pt\stroke}\phantom{Q}}}}
\def\Nmath{\vcenter{\hbox{\upright\rlap{I}\kern 1.7pt N}}}
\def\Cmath{\vcenter{\hbox{\upright\rlap{\rlap{C}\kern
                   3.8pt\stroke}\phantom{C}}}}
\def\Rmath{\vcenter{\hbox{\upright\rlap{I}\kern 1.7pt R}}}
\def\Z{\ifmmode\Zmath\else$\Zmath$\fi}
\def\Q{\ifmmode\Qmath\else$\Qmath$\fi}
\def\N{\ifmmode\Nmath\else$\Nmath$\fi}
\def\C{\ifmmode\Cmath\else$\Cmath$\fi}
\def\R{\ifmmode\Rmath\else$\Rmath$\fi}
\def\barray{\begin{eqnarray}}
\def\earray{\end{eqnarray}}
\def\beq{\begin{equation}}
\def\eeq{\end{equation}}

\def\AA{\leavevmode\setbox0=\hbox{h}
\dimen0=\ht0 \advance\dimen0 by-1ex\rlap{\raise.67\dimen0\hbox{\char'27}}A}

\makeatletter
\def\iddots{\mathinner{\mkern1mu\raise\p@
\vbox{\kern7\p@\hbox{.}}\mkern2mu
\raise4\p@\hbox{.}\mkern2mu\raise7\p@\hbox{.}\mkern1mu}}
\makeatother

\theoremstyle{plain}

\theoremstyle{remark}

\def\xi{\chi}

}

\newcommand{\kB}{k_{\mbox{\tiny{B}}}}
\newcommand{\Kcr}{K_{\mbox{\tiny{tc}}}}
\newcommand{\Dcr}{D_{\mbox{\tiny{tc}}}}

\newcommand{\eq}[1]{\begin{align}\label{#1}}
\newcommand{\en}{\end{align}}
\newcommand{\eqar}[1]{\begin{align}\label{#1}}
\newcommand{\enar}{\end{align}}

\begin{document}

\title{Form factors of the tricritical three-state Potts model in its scaling limit}

\author{Giuseppe~Mussardo\footnote{mussardo@sissa.it}}
\affiliation{SISSA and INFN, Sezione di Trieste, Via Bonomea 265, I-34136, Trieste, Italy}
\author{Marco~Panero\footnote{marco.panero@unito.it}}
\affiliation{Physics Department, University of Turin and INFN, Sezione di Torino,\\ Via Pietro Giuria 1, I-10125 Turin, Italy}
\author{Andrea~Stampiggi\footnote{astampig@sissa.it, corresponding author}}
\affiliation{SISSA and INFN, Sezione di Trieste, Via Bonomea 265, I-34136, Trieste, Italy}

\begin{abstract}

We compute the form factors of the order and disorder operators, together with those of the stress-energy tensor, of the two-dimensional three-state Potts model with vacancies along its thermal deformation of the critical point. At criticality the model is described by the non-diagonal partition function of the unitary minimal model  $\mathcal{M}_{6,7}$ of conformal field theories and is accompanied by an internal $S_3$ symmetry. Its off-critical thermal deformation is an integrable massive theory which is still invariant under $S_3$. The presence of infinitely many conserved quantities, whose spin spectrum is related to the exceptional Lie algebra $E_6$, allows us to determine the analytic $S$-matrix, the exact mass spectrum and the matrix elements of local operators of this model in an exact non-perturbative way. We use the spectral representation series of the correlators and the fast convergence of these series to compute several universal ratios of the renormalization group.

\end{abstract}

\maketitle

\clearpage
\section{Introduction}
An interesting problem in theoretical physics consists of the study of two-dimensional statistical models which exhibit a second order phase transition for certain values of their coupling constants. As well known, the proper language to frame the scaling limit of these microscopical critical models is the one of conformal field theories (CFTs) \cite{belavinInfiniteConformalSymmetry1984} which, in two dimensions, have the peculiar feature of the underlying infinite-dimensional Virasoro algebra and the possibility to define consistent models employing only a finite number of families of irreducible representations of this algebra (the so-called degenerate fields). These `minimal models' capture the universal scaling behavior of the two-dimensional critical models and are exactly solvable, in the sense that it is possible to obtain -- at least in principle --  closed expressions for the multi-point correlation functions of scaling operators, by exploiting properties of the Virasoro algebra and null-vector conditions of the corresponding degenerate fields\cite{belavinInfiniteConformalSymmetry1984,friedanConformalInvarianceUnitarity1984}. 

However, no general framework has yet been formulated for generic off-critical models. Nonetheless powerful tools of analysis can be employed if the deformed theory turns out to be integrable, i.e. if it has infinitely many commuting conserved local quantities. When this happens, these conserved quantities provide severe constraints on the structure of scattering amplitudes, which are non zero only for the elastic processes and can be computed exactly \cite{fateevCONFORMALFIELDTHEORY1990,Christe:1989ah,Christe:1989my,sotkovBootstrapFusionsTricritical1989, zamolodchikovIntegrableFieldTheory1989,Mussardo:1992uc,
mussardoStatisticalFieldTheory2020}. In practice it suffices to find all possible two-particle amplitudes, since all other scattering amplitudes involving $n>2$ particles can be expressed in terms of their ${n(n-1)/2}$ two-particle scatterings.

Notable examples of these two-dimensional integrable field theories are the ones described by the Ising model in a magnetic field (IMMF), the thermally deformed tricritical Ising model  and the thermally deformed three-state tricritical Potts model (TPM). These models possess conserved charges whose spins match the Coxeter exponents of three exceptional algebras ($E_8$, $E_7$ and $E_6$ respectively~\cite{zamolodchikovIntegrableFieldTheory1989,Christe:1989ah,Christe:1989my,sotkovBootstrapFusionsTricritical1989,Mussardo:1992uc,acerbiFormFactorsCorrelation1996,Smirnov:1992vz,mussardoStatisticalFieldTheory2020}).

Once the $S$-matrix of an integrable theory is known, it is possible to study off-critical two-point correlation functions through the technique of form factors (FFs) \cite{Karowski:1978vz,Smirnov:1992vz,mussardoStatisticalFieldTheory2020}. These quantities are defined as matrix elements of (semi-)local operators between the vacuum of the theory and multi-particle states. The novel contribution of this paper is the exact computation of FFs of the leading and sub-leading order and disorder operators in the thermal deformation of the TPM. We also compute some universal ratios of the renormalization group (see  ref.~\cite{fioravantiUniversalAmplitudeRatios2000} and references therein), which can be useful for numerical or experimental studies.

The $q$-state Potts model has been studied in great detail in the literature - see e.g. ref.~\cite{wuPottsModel1982} for a review. A generalization of this model, called `dilute' due to the presence of vacancies, has been studied in
\cite{nienhuisFirstSecondOrderPhase1979}. Its lattice realization has a Hamiltonian, invariant under the permutation group $S_3$, of the form
\begin{align}
\label{eq_diluted_potts_nienhuis}
\mathcal{H} = -\sum_{\langle i,j\rangle} t_i t_j\left( \mathcal{K} + \mathcal{F} \delta_{s_i, s_j}\right) + \mathcal{D} \sum_i t_i \,\,\,,
\end{align}
where each site $i$ is associated with a Potts `spin' ${s_i = \{0,1,2\}}$ and a vacancy variable ${t_i = \{0,1\}}$. In the latter equation $\mathcal{K}$ and $\mathcal{F}$ are interaction parameters associated to vacancies and Potts spins respectively, while $\mathcal{D}$ plays the role of chemical potential associated to vacancies. The usual Potts model is retrieved when the chemical potential $\mathcal{D}$ is negative and large in absolute value. Since the TPM possesses only two relevant $S_3$ invariant operators \cite{friedanConformalInvarianceUnitarity1984}, the energy $\epsilon$ and the vacancy $t$, it is retrieved from eq.~\eqref{eq_diluted_potts_nienhuis} in the limit of vanishing $\mathcal{K}$. The TPM is also exactly solvable in a honeycomb lattice and proven to be self-dual, i.e. there is a mapping between its disordered (high-temperature) and ordered (low-temperature) phases \cite{nienhuisLocusTricriticalTransition1991}.

The content of this paper is organised as follows. In Section \ref{s1} we recall a few basic properties of minimal models of CFT, useful to set mostly the notation, followed by a thorough discussion of the TPM, in particular the operator content and the fusion rules of this model. Section \ref{s2} is dedicated to the thermal deformation of the TPM, regarded as an integrable field theory; the scattering theory and the mass spectrum are discussed there. The FFs are introduced in Section \ref{s3} (their analytic properties are gathered in Appendix \ref{a1}). In Section \ref{s5} we review the FFs of the stress-energy tensor, originally obtained in \cite{acerbiFormFactorsCorrelation1996}. The FFs of the order and disorder operators of the thermal deformation of the TPM are presented in Section \ref{s4}. The two-point correlation functions and the universal ratios of the renormalization group are discussed in Section \ref{s6}. Some preliminary results coming from the Monte Carlo simulations of the model are presented in Section  
\ref{sec:comparison_with_a_Monte_Carlo_study}. Our conclusions are finally summarised in Section \ref{sec:discussion_and_conclusions}.

%%%%%%%%%%

\section{Critical TPM: the CFT Approach}\label{s1}
As well known, minimal models $\mathcal{M}_{p,q}$ of CFT are characterized by the following properties \cite{belavinInfiniteConformalSymmetry1984, friedanConformalInvarianceUnitarity1984}:
\begin{enumerate}[i)]
\item their central charge $c$ is always less than one and can be parameterized by two coprime integers $p, q$ (with $q > p$)
\begin{equation}\label{central_charge}
c(p,q) = 1- 6\frac{(p-q)^2}{pq}.
\end{equation}
\item there is a finite number of families of degenerate fields, both in the analytic and anti-analytic sector of the CFT.  In the analytic sector, a degenerate primary field $\phi_{r,s}(z)$ has conformal dimension
\begin{equation}
\label{conformal_dimension_degenerate_fields}
\Delta_{r,s}= \frac{(pr-qs)^2-(q-p)^2}{4pq},
\end{equation}
with $1\leq r\leq q-1$ and $1\leq s\leq p-1$, coincident to that of the anti-analytic counterpart $\overline{\phi}_{r,s}(\overline{{z}})$: $\overline{\Delta}_{r,s} = \Delta_{r,s}$.
Unitary minimal models are retrieved by setting in the previous formulas $m=q=p+1$ with $m\geq 3$.
\item correlation functions of all conformal fields satisfy a set of linear differential equations that can be exactly solved, yielding in turn the structure constants of the operator algebra.
\item the modular-invariant partition function can be determined exactly \cite{cappelliModularInvariantPartition1987, cardyEffectBoundaryConditions1986, cardyOperatorContentTwodimensional1986}.
\end{enumerate}

Another important feature of minimal models is that the operator product expansion (OPE) algebra of degenerate fields, in the analytic and anti-analytic sector separately, is `closed': the OPE of two degenerate fields can be written in terms of degenerate fields and their descendants. In particular, the operators appearing in the OPE are obtained by iterative multiplication of the fields $\phi_{2,1}$ and $\phi_{1,2}$. The general formula, which does not give information about the structure constants, is
\begin{equation}
\label{fusion_rules}
\phi_{r_1,s_1}\times\phi_{r_2, s_2} = \sum_{r_3= \{\abs{r_2-r_1}+1, \abs{r_2-r_1}+3, \dots\}}^{\min(r_1+r_2-1, 2q-1-r_1-r_2)}\sum_{s_3= \{\abs{s_2-s_1}+1,\abs{s_2-s_1}+3, \dots\}}^{\min(s_1+s_2-1, 2p-1-s_1-s_2)} \phi_{r_3, s_3}.
\end{equation}

Finding the modular-invariant partition functions of a given universality class associated to a minimal model is equivalent to determining the operator content of the model. The problem of the modular-invariant partition functions for unitary theories was first addressed by Cardy \cite{cardyEffectBoundaryConditions1986, cardyOperatorContentTwodimensional1986}, and the solution for minimal models with periodic boundary conditions was found soon after by Cappelli {\it et al.}~\cite{cappelliModularInvariantPartition1987}.

The conformal dimensions of the fields in $\mathcal{M}_{6,7}$ are summarized in Table \ref{TPM_weights}. The universality class $\mathcal{M}_{6,7}$ contains two algebras of fields which close under the OPE: the pentacritical $\varphi^{10}$ 
Landau-Ginzburg field theory and the TPM. 
\begin{table}
\centering
\def\arraystretch{1.5}
\begin{tabular}{|l || c c c c c c|} 
 \hline
 5 & $5$ & $\frac{22}{7}$ & $\frac{12}{7}$ & $\frac{5}{7}$ & $\frac{1}{7}$ & $0$\\
 4 & $\frac{23}{8}$ & $\frac{85}{56}$ & $\frac{33}{56}$ & $\frac{5}{56}$ & $\frac{1}{56}$ & $\frac{3}{8} $\\
 3 & $\frac{4}{3}$ & $\frac{10}{21}$ & $\frac{1}{21}$ & $\frac{1}{21}$ & $\frac{10}{21}$ & $\frac{4}{3} $\\ 
 2 & $\frac{3}{8}$ & $\frac{1}{56}$ & $\frac{5}{56}$ & $\frac{33}{56}$ & $\frac{85}{56}$ & $\frac{23}{8} $\\ 
 1 & $0$ & $\frac{1}{7}$ & $\frac{5}{7}$ & $\frac{12}{7}$ & $\frac{22}{7}$ & $5$\\
 \hline\hline
 s $\uparrow$, r $\rightarrow$ & 1 & 2 & 3 & 4 & 5 & 6\\
 \hline
\end{tabular}
\caption{Ka\v c table of the universality class $\mathcal{M}_{6,7}$.}
\label{TPM_weights}
\end{table}
In correspondence to the minimal model $\mathcal{M}_{6,7}$ there are then two modular-invariant partition functions: a diagonal one, which corresponds to the $\varphi^{10}$ Landau-Ginzburg theory,
\begin{equation}
\label{M67_diagonal_partition}
\begin{aligned}
Z^{(1)}&= \abs{\chi_{0}}^2+\abs{\chi_{\frac{1}{7}}}^2 + \abs{\chi_{\frac{5}{7}}}^2 + \abs{\chi_{\frac{12}{7}}}^2 + \abs{\chi_{\frac{22}{7}}}^2
 +  \abs{\chi_{5}}^2 +  \abs{\chi_{\frac{3}{8}}}^2+\abs{\chi_{\frac{1}{56}}}^2+\abs{\chi_{\frac{5}{56}}}^2+\\
&\quad + \abs{\chi_{\frac{33}{56}}}^2
 +\abs{\chi_{\frac{85}{56}}}^2+ \abs{\chi_{\frac{23}{8}}}^2 +
\abs{\chi_{\frac{4}{3}}}^2+ \abs{\chi_{\frac{10}{21}}}^2 +\abs{\chi_{\frac{1}{21}}}^2
\end{aligned}
\end{equation}
and an off-diagonal one relative to the TPM 
\begin{equation}
\label{M67_off_diagonal_partition}
%\begin{aligned}
Z^{(2)}
%&
= \abs{\chi_{0}+\chi_{5}}^2 + \abs{\chi_{\frac{1}{7}}+\chi_{\frac{22}{7}}}^2 +\abs{\chi_{\frac{5}{7}}+\chi_{\frac{12}{7}}}^2 
%\\
%&\quad 
+ 2\abs{\chi_{\frac{4}{3}}}^2+2\abs{\chi_{\frac{10}{21}}}^2+2\abs{\chi_{\frac{1}{21}}}^2.
%\end{aligned}
\end{equation}
Here $\chi_{\Delta}$ is the character of the irreducible representation of the Virasoro algebra of weight $\Delta$, whose explicit formulas can be found in \cite{rocha-caridiCharactersIrreducibleRepresentations1984}.

Using the notation of Table \ref{t_operator_content_TPM} for the various fields, it is possible to show that the fusion rules of Table \ref{t_fusion_TPM} are compatible with a $S_3$ internal symmetry, so that operators can be organized according to irreducible representations of this discrete group, i.e. either in doublets ($\vb{s}$, $\vb{Z}$ and $\vb{A}$) or in  the one-dimensional neutral representation ($\mathbbm{1}$, $\epsilon$, $t$, $\vb{X}$, $\vb{Y}$ and $\vb{B}$). In the model there are no operators which transform according to the alternating representation.
\begin{table}
\centering
\def\arraystretch{1.5}
\makebox[\linewidth]{
\begin{tabular}{|c c c c c|}
\hline
$\phi_{r,s}$ & $(\Delta,\bar{\Delta})$ & Operator  &Interpretation & $S_3$ representation\\
\hline\hline
$\phi_{1,1}\times \bar{\phi}_{1,1}$ & $\left(0,0\right)$ & $\mathbbm{1}$ & Identity & $1$ \\
$\phi_{3,3}\times \bar{\phi}_{3,3}$ & $\left(\frac{1}{21},\frac{1}{21}\right)$ & $\vb{s}$ & Leading Order & $2$\\
$\phi_{2,1}\times \bar{\phi}_{2,1}$ & $\left(\frac{1}{7},\frac{1}{7}\right)$ & $\epsilon$ & Energy & $1$\\
$\phi_{2,3}\times \bar{\phi}_{2,3}$ & $\left(\frac{10}{21},\frac{10}{21}\right)$ & $\vb{Z}$ & Subleading Order & $2$\\
$\phi_{3,1}\times \bar{\phi}_{3,1}$ & $\left(\frac{5}{7},\frac{5}{7}\right)$ & $t$ & Vacancy & $1$\\
$\phi_{1,3}\times \bar{\phi}_{1,3}$ & $\left(\frac{4}{3},\frac{4}{3}\right)$ & $\vb{A}$ & Irrelevant & $2$\\
$\phi_{4,1}\times \bar{\phi}_{4,1}$ & $\left(\frac{12}{7},\frac{12}{7}\right)$ & $\vb{X}$ & Irrelevant & $1$\\
$\phi_{5,1}\times \bar{\phi}_{5,1}$ & $\left(\frac{22}{7},\frac{22}{7}\right)$ & $\vb{B}$ & Irrelevant & $1$\\
$\phi_{6,1}\times \bar{\phi}_{6,1}$ & $\left(5,5\right)$ & $\vb{Y}$  & Irrelevant & $1$\\
$\phi_{4,1}\times \bar{\phi}_{3,1}$ & $\left(\frac{12}{7},\frac{5}{7}\right)$ & $\mathcal{F}$ & Irrelevant, Spin 1 & $1$\\
$\phi_{3,1}\times \bar{\phi}_{4,1}$ & $\left(\frac{5}{7},\frac{12}{7}\right)$ & $\bar{\mathcal{F}}$ & Irrelevant, Spin -1 & $1$\\
$\phi_{4,1}\times \bar{\phi}_{2,1}$ & $\left(\frac{12}{7},\frac{1}{7}\right)$ & $\mathcal{U}$ & Irrelevant, Spin 3 & $1$\\
$\phi_{2,1}\times \bar{\phi}_{4,1}$ & $\left(\frac{1}{7},\frac{12}{7}\right)$ & $\bar{\mathcal{U}}$ & Irrelevant, Spin -3 & $1$\\
$\phi_{6,1}\times \bar{\phi}_{1,1}$ & $\left(5,0\right)$ & $\mathcal{W}$  & Irrelevant, Spin 5 & $1$\\
$\phi_{1,1}\times \bar{\phi}_{6,1}$ & $\left(0,5\right)$ & $\bar{\mathcal{W}}$ & Irrelevant, Spin -5 & $1$\\
\hline
\end{tabular}
}
\caption{Operator content of the TPM at criticality.}
\label{t_operator_content_TPM}
\end{table}
\begin{table}
\centering
\def\arraystretch{1.5}
\makebox[\linewidth]{
\begin{tabular}{|c l l| c l l|}
\hline
$\vb{s} \times \vb{s}$ & $=$ & $\mathbbm{1}+\epsilon +t +\vb{X}+\vb{B}+\vb{Y}+\vb{s}+\vb{Z}+\vb{A}$ & 
$\epsilon \times \epsilon$ & $=$ & $\mathbbm{1}+t$\\
$\vb{Z} \times \vb{Z}$ & $=$ & $\mathbbm{1}+\epsilon +t +\vb{X}+\vb{Y}+\vb{s}+\vb{A}$ & 
$t \times t$ & $=$ & $\mathbbm{1}+t+\vb{B}$\\
$\vb{A} \times \vb{A}$ & $=$ & $\mathbbm{1}+\vb{Y}+\vb{A}$ & $\vb{X} \times \vb{X}$ & $=$ & $\mathbbm{1}+t+\vb{B}$\\
$\vb{B} \times \vb{B}$ & $=$ & $\mathbbm{1}+t$ & $\vb{Y} \times \vb{Y}$ & $=$ & $\mathbbm{1}$\\
\hline
$\vb{s} \times \epsilon$ & $=$ & $\vb{s}+\vb{Z}$ & 
$\vb{s} \times t$ & $=$ & $\vb{s}+\vb{Z}+\vb{A}$ \\
$\vb{s} \times \vb{Z}$ & $=$ & $\epsilon +t +\vb{X}+\vb{B}+\vb{s}+\vb{Z}$ & 
$\vb{s} \times \vb{A}$ & $=$ & $t +\vb{X}+\vb{s}$\\
$\vb{s} \times \vb{X}$ & $=$ & $\vb{s}+\vb{Z}+\vb{A}$ & 
$\vb{s} \times \vb{B}$ & $=$ & $\vb{s} +\vb{Z}$\\
$\vb{s} \times \vb{Y}$ & $=$ & $\vb{s}$ & 
 & & \\
\hline
$\epsilon \times \vb{Z}$ & $=$ & $\vb{s}+\vb{A}$ & 
$\epsilon \times t$ & $=$ & $\epsilon+\vb{X}$\\
$\epsilon \times \vb{A}$ & $=$ & $\vb{Z}$ & 
$\epsilon \times \vb{X}$ & $=$ & $t+\vb{Y}$\\
$\epsilon \times \vb{B}$ & $=$ & $\vb{X}+\vb{Y}$ & 
$\epsilon \times \vb{Y}$ & $=$ & $\vb{B}$\\
\hline
$\vb{Z} \times t$ & $=$ & $\vb{s}+\vb{Z}$ &
$\vb{Z} \times \vb{A}$ & $=$ & $\epsilon+\vb{B}+\vb{Z}$\\
$\vb{Z} \times \vb{X}$ & $=$ & $\vb{s}+\vb{Z}$ &
$\vb{Z} \times \vb{B}$ & $=$ & $\vb{s}+\vb{A}$\\
$\vb{Z} \times \vb{Y}$ & $=$ & $\vb{Z}$ & 
 & & \\
\hline
$t \times \vb{A}$ & $=$ & $\vb{s}$ & 
$t \times \vb{X}$ & $=$ & $\epsilon+\vb{X}+\vb{Y}$\\
$t \times \vb{B}$ & $=$ & $t+\vb{B}$ & 
$t \times \vb{Y}$ & $=$ & $\vb{X}$\\
\hline
$\vb{A} \times \vb{X}$ & $=$ & $\vb{s}$ &
$\vb{A} \times \vb{B}$ & $=$ & $\vb{Z}$\\
$\vb{A} \times \vb{Y}$ & $=$ & $\vb{A}$ & 
 & & \\
\hline
$\vb{X} \times \vb{B}$ & $=$ & $\epsilon+\vb{X}$ & 
$\vb{X} \times \vb{Y}$ & $=$ & $t$\\
\hline
$\vb{B} \times \vb{Y}$ & $=$ & $\epsilon$ & 
 & & \\
 \hline
\end{tabular}
}
\caption{Fusion rules of the TPM at criticality.}
\label{t_fusion_TPM}
\end{table}

Doublet operators like $\vb{s}$ and $\vb{Z}$ are charged under the generator $\Omega$ of the subgroup $Z_3$. Let the components of the doublet corresponding to the leading operator $\vb{s}$ be respectively $\sigma$ and $\tilde{\sigma}$. Similarly one defines the components of the subleading order operator $\vb{Z}$ to be $Z$ and $\tilde{Z}$. As it happens in the case of the Ising model \cite{fradkinDisorderOperatorsTheir2017}, the tricritical Ising model \cite{cortescuberoDualityFormFactors2022}
and the three-state Potts model \cite{capizziEntanglementStatePotts2021, casellePottsCorrelatorsStatic2006}, in view of the self-duality of the lattice model we can introduce a leading disorder operator ${(\mu,\tilde{\mu})}$ associated to the leading order operator ${(\sigma,\tilde{\sigma})}$. In complete analogy to the leading counterpart, ${(\zeta,\tilde{\zeta})}$ is the subleading disorder operator associated to the subleading order operator ${(Z,\tilde{Z})}$. The conformal dimensions of these operators are:
\begin{equation}
\label{eq_scaling_dimensions_magnetization}
\begin{aligned}
&\left(\Delta_{\sigma},\bar{\Delta}_{\sigma}\right)=
\left(\Delta_{\tilde{\sigma}},\bar{\Delta}_{\tilde{\sigma}}\right)=
\left(\Delta_{\mu},\bar{\Delta}_{\mu}\right)=
\left(\Delta_{\tilde{\mu}},\bar{\Delta}_{\tilde{\mu}}\right)=
\left(\frac{1}{21},\frac{1}{21}\right),\\
&\left(\Delta_{Z},\bar{\Delta}_{Z}\right)=
\left(\Delta_{\tilde{Z}},\bar{\Delta}_{\tilde{Z}}\right)=
\left(\Delta_{\zeta},\bar{\Delta}_{\zeta}\right)=
\left(\Delta_{\tilde{\zeta}},\bar{\Delta}_{\tilde{\zeta}}\right)=
\left(\frac{10}{21},\frac{10}{21}\right).
\end{aligned}
\end{equation}

The transformation properties under $S_3$ of the order and disorder operators are as follows. Under the generator $\Omega$ of $Z_3$ the disorder operators are invariant and the order operators acquire a phase:
\begin{equation}
\label{eq_z3_charge_magnetization}
\begin{aligned}
&\Omega \sigma = \omega \sigma,  & & 
\Omega \tilde{\sigma} = \omega^* \tilde{\sigma}, & & 
\Omega Z = \omega^2 Z,  & &
\Omega \tilde{Z} = (\omega^*)^2 \tilde{Z},\\
&\Omega \mu = \mu, & & 
\Omega\tilde{\mu} = \tilde{\mu}, & & 
\Omega \zeta = \zeta, & & 
\Omega\tilde{\zeta} = \tilde{\zeta},
\end{aligned}
\end{equation}

where $\omega = \exp(2i\pi /3)$. On the other hand, the generator $\mathcal{C}$ of $Z_2$ acts like a charge conjugation operator:
\begin{equation}
\label{eq_z2_charge_conj_magnetization}
\begin{aligned}
\mathcal{C}\sigma &= \tilde{\sigma}, & & &\mathcal{C}Z &=\tilde{Z},\\
\mathcal{C}\mu &= \tilde{\mu}, & & &\mathcal{C}\zeta &=\tilde{\zeta}.
\end{aligned}
\end{equation}

The assumption that only order operators are charged with respect to $Z_3$ is consistent with an alternative definition of TPM, in terms of a parafermionic algebra of central charge $6/7$ -- see eq.~(A.8) of ref.~\cite{zamolodchikovNonlocalParafermionCurrents1985}. The parafermionic theory is defined to be invariant under the product group $Z_3\times \tilde{Z}_3$. The first $Z_3$ group acts nontrivially only on order operators, while the second only on disorder ones. However, the symmetry group of the TPM is $S_3$, which contains $Z_3$ and $Z_2$ as subgroups. While $Z_3$ is the group of symmetry of the parafermionic theory, the second group $Z_2$ acts as a charge conjugation operator. 

There are two important features of parafermionic theories. The first is that they are self-dual by definition: order and disorder operators in the ordered phase can be mapped into disorder and order operators of the disordered phase respectively. As a consequence, we will only focus on the FFs in the disordered phase. The second is that order and disorder operators are not mutually local, in general. In particular, for the TPM the non-locality factors are 
\begin{equation}
\label{eq_nonlocality_factor}
\begin{aligned}
&\gamma_{\sigma,\mu} = \gamma_{Z,\zeta} = -1/3, & &
\gamma_{\sigma,\zeta} = \gamma_{Z,\mu} = +1/3,\\
&\gamma_{\tilde{\sigma},\mu} = \gamma_{\tilde{Z},\zeta} = 1/3, & &
\gamma_{\tilde{\sigma},\zeta} = \gamma_{\tilde{Z},\mu} = -1/3.
\end{aligned}
\end{equation}
Identical results apply if we charge-conjugate both the first and second entry in the $\gamma$'s.

%%%%%%%%%%%%%

\section{Thermal deformation of the TPM} \label{s2}
Consider the fixed point action $\mathcal{A}_0$ of the TPM  and define the action $\mathcal{A}$ of its thermal deformation as
\begin{equation}\label{eq_thermally_deformed_TPM_action}
\mathcal{A} = \mathcal{A}_{0}+\lambda\int \dd^2 x\ \epsilon(x),
\end{equation}
where $\epsilon$ is the energy operator and $\lambda$ is its coupling constant. In the scaling limit of the lattice version of the model, i.e. the limit in which the temperature $T$ approaches its critical value $T_c$ and the chemical potential is fixed at its critical value, we have $\lambda \propto (T-T_c)$. The action ${\mathcal A}$ corresponds to a massive theory  and there is a non-vanishing trace of the stress-energy tensor $\Theta(z,\bar{z})$ given by  
\begin{equation}\label{eq_trace_stress_deformation}
\Theta(z,\bar{z}) = 2\pi \lambda (2-2\Delta_\epsilon)\epsilon(z,\bar{z}).
\end{equation} 
In the following we will often denote the quantum field theory associated to the thermal deformation of the TPM as $\epsilon$-TPM. 
In addition to the stress-energy tensor, in the $\epsilon$-TPM there are infinitely many local conserved quantities which commute with each other and therefore the theory is integrable~\cite{fateevCONFORMALFIELDTHEORY1990, zamolodchikovIntegrableFieldTheory1989}. The spins of conserved quantities for this model are given by the infinite sequence 
of residues 
\begin{equation}
\{1, 4, 5, 7, 8,11\} \hspace{3mm} {\rm mod \,\, h=12 }\,\,\,.
\end{equation}
In addition, they coincide, modulo the Coxeter number $h=12$, with the Coxeter exponents of the exceptional algebra $E_6$.  

\comment{Peculiar to the $\epsilon$-TPM is the presence of even-spin conserved quantities. While odd-spin conserved quantities in the three models come from the identity family, even-spin conserved quantities in the $\epsilon$-TPM come from the spin-5 field $\mathcal{W}$, defined in Table \ref{t_operator_content_TPM}.}

\begin{figure}
    \centering
    \makebox[\textwidth][c]{
\begin{tikzpicture}
    \node [circle,fill=black,inner sep=0pt,minimum size=5pt](o) at (0,0){};
    \node [ circle,fill=black,inner sep=0pt,minimum size=5pt](l) at (-1,0){};
    \node [ circle,fill=black,inner sep=0pt,minimum size=5pt](ll) at (-2,0){};
    \node [ circle,fill=black,inner sep=0pt,minimum size=5pt](r) at (1,0){};
    \node [ circle,fill=black,inner sep=0pt,minimum size=5pt](rr) at (2,0){};
    \node [ circle,fill=black,inner sep=0pt,minimum size=5pt](u) at (0,1){};
    \node [ circle,fill=black,inner sep=0pt,minimum size=5pt](uu) at (0,2){};
    \draw[] (ll)--(rr);
    \draw[] (o)--(u);
    \draw[] (u)--(uu);
    \node[]() at (0,-0.5){$(3;3)$};
    \node[]() at (-1,-0.5){$(2;2)$};
    \node[]() at (-2,-0.5){$(1;1)$};
    \node[]() at (1,-0.5){$(4;2)$};
    \node[]() at (2,-0.5){$(5;1)$};
    \node[]() at (0.5,1){$(6;2)$};
    \node[]() at (0.5,2){$(0;1)$};
\end{tikzpicture}
}
    \caption{Dynkin diagram of the affine $E_6$ algebra. The first term in parenthesis labels the root, while the second its comark.}
    \label{f_e6aff}
\end{figure}
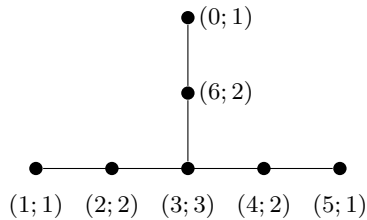

The hidden $E_6$ structure of the model can be understood employing a different CFT description of the TPM, i.e. the so-called coset construction \cite{difrancescoConformalFieldTheory1997}, which in this case consists in $[(E_6)_1\times(E_6)_1]/(E_6)_2$. Let us discuss this point in more detail. The affine algebra $(E_6)_k$ (Fig.~\ref{f_e6aff}) has different integrable highest weights depending on the level $k$. Calling $w_i$ ($i=0,1,\ldots, 7$) the fundamental weights of $(E_6)_k$, which are the basis of the weight space in the natural basis $w_i = \hat{e}_{i+1}$ (e.g. $w_1 = (0,1,0,0,0,0,0)$), the integrable highest weights at level $k=1$ are $w_0$, $w_1$ and $w_5$, while at level $k=2$ we have $2w_0$, $2w_1$, $2w_5$, $(w_0+w_1)$, $(w_0+w_5)$, $(w_1+w_5)$, $w_2$ and $w_6$. Through the coset fields, we can recover the operator content of the TPM. As discussed for instance in \cite{difrancescoConformalFieldTheory1997}, the procedure consists of finding the allowed branchings, the fractional dimensions of the integrable representations, conjugate representations (related to the symmetry group of the non-affine Dynkin diagram) and coset fields, i.e. classes of equivalent branchings related by automorphisms ($Z_3$ in the case of affine $E_6$). 
%Even though the mapping from coset fields to Virasoro operators is not one-to-one, conformal dimensions neatly match those of the TPM. 
The results are summarized in Table \ref{t_coset}.

\begin{table}[b]
\centering
\def\arraystretch{1.5}
\makebox[\linewidth]{
\begin{tabular}{|l| c |}
\hline
Cosets & Operators \\
\hline
$\lbrace
\left[w_0,w_0,2w_0\right]\sim
\left[w_1,w_1,2w_5\right]\sim
\left[w_5,w_5,2w_5\right]
\rbrace$
& $\mathbbm{1}, \vb{Y}$ \\
$\lbrace
\left[w_0,w_1,w_0+w_1\right]\sim 
\left[w_5,w_0,w_0+w_5\right]\sim
\left[w_1,w_5,w_1+w_5\right]
\rbrace$
& $\sigma$ \\
$\lbrace
\left[w_1,w_0,w_0+w_1\right]\sim 
\left[w_0,w_5,w_0+w_5\right]\sim
\left[w_5,w_1,w_1+w_5\right]
\rbrace$
& $\tilde{\sigma}$ \\
$\lbrace
\left[w_0,w_0,w_6\right]\sim
\left[w_1,w_1,w_2\right]
\rbrace$
& $\epsilon, \vb{B}$ \\
$\lbrace
\left[w_1,w_5,w_6\right]\sim
\left[w_5,w_0,w_2\right]
\rbrace$
& $Z$ \\
$\lbrace
\left[w_5,w_0,w_2\right]\sim
\left[w_5,w_1,w_6\right]
\rbrace$
& $\tilde{Z}$ \\
$\lbrace
\left[w_0,w_0,w_1+w_5\right]\sim
\left[w_1,w_1,w_0+w_5\right]\sim
\left[w_5,w_5,w_0+w_1\right]
\rbrace$
& $t, \vb{X}$ \\
$\lbrace
\left[w_0,w_1,2w_5\right]\sim 
\left[w_5,w_0,2w_1\right]\sim
\left[w_1,w_5,2w_0\right]
\rbrace$
& $A$ \\
$\lbrace
\left[w_1,w_0,2w_5\right]\sim 
\left[w_0,w_5,2w_1\right]\sim
\left[w_5,w_1,2w_0\right]
\rbrace$
& $\tilde{A}$ \\
\hline
\end{tabular}
}
\caption{Identification of coset fields of $[(E_6)_1\times(E_6)_1]/(E_6)_2$ with the operator content of the TPM. An element of the coset $[\lambda,\mu,\nu]$ is such that it appears in the branching $\lambda + \mu = \sum_{\nu}b_{\lambda,\mu}^\nu \nu$.}
\label{t_coset}
\end{table}
The coset construction of the TPM only holds for the critical theory. Away from criticality there is however a correspondence between the thermal deformation of the TPM and the  exceptional algebra $E_6$ and this involves the Toda field theories. These theories are built starting from a Lagrangian associated to a Lie algebra of finite rank (see, e.g. \cite{mussardoStatisticalFieldTheory2020, Braden:1989bu} and references therein). In \cite{eguchiDeformationsConformalField1989} it was proven that the deformation of the Toda field theory of $E_6$ perturbed with the adjoint representation is integrable and has conserved charges whose spins coincide with those of the thermal deformation of the TPM. Thus the scattering matrix of this perturbed Toda field theory has the same minimal scattering matrix as our model: its mass spectrum contains six particles, two particle-antiparticle doublets, $(l,\bar{l})$ and $(h, \bar{h})$, and two self-conjugated particles, $L$ and $H$, and the masses of these particles can be expressed as
\begin{equation}\label{eq_masses_TPM}
\begin{aligned}
    & m_l = m_{\bar{l}} = M(\lambda ), & &
    m_L = 2\cos\left(\pi/4\right)m_l,\\
    & m_h = m_{\bar{h}} = 2\cos\left(\pi/12\right)m_l, & &
    m_H = 4[\cos\left(\pi/4\right)]^2 m_l,
\end{aligned} 
\end{equation}
where $M(\lambda)$ is the mass of the lightest kink of the theory and depends on the coupling constant $\lambda$ in eq.~\eqref{eq_thermally_deformed_TPM_action} as \cite{fateevExactRelationsCoupling1994}
\begin{equation}
    \label{eq_mass_lightestkink}
    M(\lambda) = C_{\epsilon} |\lambda|^{\frac{7}{12}},
\end{equation}
where
\begin{equation}\label{eq_mass_constant}
    C_{\epsilon} = \left[4\pi^2\lambda^2
    \gamma\left(\frac{4}{7}\right)\gamma\left(\frac{9}{14}\right)\gamma\left(\frac{5}{7}\right)\gamma\left(\frac{11}{14}\right)
    \right]^{\frac{7}{24}}\frac{2\Gamma\left(\frac{1}{4}\right)\Gamma\left(\frac{13}{12}\right)}{\Gamma\left(\frac{1}{2}\right)\Gamma\left(\frac{2}{3}\right)\Gamma\left(\frac{7}{6}\right)}
\approx 3.74656.
\end{equation}
In the latter equation $\gamma(z)$ is a shorthand for $\Gamma(z)/\Gamma(1-z)$. 
\comment{The important property of these six masses is that the vector that can be built starting from them is nothing more than the Perron-Frobenius eigenvector of the incidence matrix $I=2-C$, where $C$ is the Cartan matrix of $E_6$. The implication of this statement is that masses can be put into correspondence with the Dynkin diagram of $E_6$, as in Figure~\ref{f_masses}.  
\begin{figure}
    \centering
    \makebox[\textwidth][c]{
\begin{tikzpicture}
    \node [circle,fill=black,inner sep=0pt,minimum size=5pt](o) at (0,0){};
    \node [ circle,fill=black,inner sep=0pt,minimum size=5pt](l) at (-1,0){};
    \node [ circle,fill=black,inner sep=0pt,minimum size=5pt](ll) at (-2,0){};
    \node [ circle,fill=black,inner sep=0pt,minimum size=5pt](r) at (1,0){};
    \node [ circle,fill=black,inner sep=0pt,minimum size=5pt](rr) at (2,0){};
    \node [ circle,fill=black,inner sep=0pt,minimum size=5pt](u) at (0,1){};
    \draw[] (ll)--(rr);
    \draw[] (o)--(u);
    \node[]() at (0,-0.5){$m_H$};
    \node[]() at (-1,-0.5){$m_{h}$};
    \node[]() at (-2,-0.5){$m_{l}$};
    \node[]() at (1,-0.5){$m_{\bar{h}}$};
    \node[]() at (2,-0.5){$m_{\bar{l}}$};
    \node[]() at (0,1.5){$m_{L}$};
\end{tikzpicture}
}\caption{Relation between the Dynkin diagram of $E_6$ and the six masses of the thermally deformed bidimensional three-state tricritical Potts model.}
    \label{f_masses}
\end{figure}
}

For the thermal deformation of the TPM, the excitations are ordinary particles only in the disordered phase $\lambda>0$, while some of them become kinks in the ordered phase $\lambda<0$. In a quantum field theory with localized interactions, particles are associated to asymptotic states which transform according to irreducible representations of the symmetry group of the theory. For the thermal deformation of the TPM, we have two $S_3$-doublets and two singlets. Since the particles in the doublets $(l,\bar{l})$ and $(h,\bar{h})$ are generated by order (magnetization-like) operators, we define the non-locality properties of the particles with respect to disorder operators to be
 \begin{equation}
\label{eq_nonlocality_l}
\begin{aligned}
&\gamma_{l,\mu} = \gamma_{l,\zeta} = 1/3, & & \gamma_{\bar{l},\mu} = \gamma_{\bar{l},\zeta} = -1/3.
\end{aligned}
\end{equation}
Identical relations hold if $l$ is replaced by $h$. This choice of non-local factors is consistent with the scattering theory of the $\epsilon$-TPM, which will now be outlined.

\subsection{Scattering Theory}
In ${(1+1)}$-dimensional theories, all particles scatter along a straight line.  It is useful to use the rapidity variable $\theta$ 
which parameterizes the on-shell momentum of the particle of type $a$ and mass $m_a$ as 
\begin{equation}
\label{eq_rapidity}
p_a(\theta_a) = m_a (\cosh \theta_a, \sinh \theta_a)
\end{equation}
and denote the one-particle state as $|A_a(\theta)\rangle$. For multiparticle states we assume, as a convention, that all incoming particles can be written from left to right, with decreasing rapidity. On the other hand, outgoing states are written in the converse: particles with the highest rapidity on the right and particles with lowest rapidity on the left. Examples are provided in Table \ref{t_convention}.
\begin{table}
\centering
\def\arraystretch{1.5}
\makebox[\linewidth]{
\begin{tabular}{|c| c | c|}
\hline
Picture & State & Convention\\
\hline
\multirow{2}{*}{\begin{tikzpicture}[baseline=0]
  \node [ ](xmax) at (2,0){};
  \node [ ](xmin) at (-2.5,0){};
  \node [ ](lab)  at (2,-0.3){$\theta$};
  \node [circle,fill=black,inner sep=0pt,minimum size=5pt ](A1)   at (-2, 0){};
 \node[ ](Aname1) at (-2, 0.3) {$A_a(\theta_1)$};
    \node [circle,fill=black,inner sep=0pt,minimum size=5pt ](A2)   at (-1, -0){};
    \node[ ](Aname2) at (-1, -0.3) {$A_b(\theta_2)$};
        \node [circle,fill=black,inner sep=0pt,minimum size=5pt ](A3)   at (0, 0){};
        \node[ ](Aname3) at (0, 0.3) {$A_c(\theta_3)$};
        \node [circle,fill=black,inner sep=0pt,minimum size=5pt ](A4)   at (1.5, 0){};
        \node[ ](Aname4) at (0.75, -0.3) {$\dots$};
  %%%
  \draw [-stealth] (xmin)--(xmax);
\end{tikzpicture}} & Incoming & $\displaystyle{|A_a(\theta_1)A_b(\theta_2)A_c(\theta_3)\dots \rangle_{IN}}$\\ 
& Outgoing & $\displaystyle{|\dots A_c(\theta_3)A_b(\theta_2)A_a(\theta_1)\rangle_{OUT}}$\\
\hline
\end{tabular}
}
\caption{Conventions for incoming and outgoing states, where $\theta_1 > \theta_2 > \theta_3$.}
\label{t_convention}
\end{table}

Integrable models have commuting conserved quantities $Q_s$ that can be diagonalized simultaneously with the Hamiltonian. This condition imposes strong constraints on the scattering matrix. Apart from being unitary and satisfying crossing symmetry, the $S$-matrix is elastic and all $n$-particle processes can be factorized in terms of two-particle scatterings \cite{zamolodchikovIntegrableFieldTheory1989,mussardoStatisticalFieldTheory2020}. These two properties, along with the assumptions of bootstrap, i.e. treating bound states on the same footing as asymptotic particles, allow for the determination of the analytic two-body scattering amplitudes.
In the case of the $\epsilon$-TPM, the $S$-matrix \cite{fateevCONFORMALFIELDTHEORY1990, sotkovBootstrapFusionsTricritical1989} is diagonal, i.e. reflection amplitudes
vanish, and the various amplitudes are reported in Table \ref{t_smatrix_TPM}. They can be expressed in terms of 
the building block functions  
\begin{equation}
[x] \equiv s_{x/12}(\theta) = \frac{\sinh [(\theta+i\pi x/12)/2]}{\sinh [(\theta-i\pi x/12)/2]}\,\,\,.
\label{buildingblockk}
\end{equation}
\begin{table}
\centering
\def\arraystretch{2}
\makebox[\textwidth][c]{
\begin{tabular}{|m{0.03\textwidth}<{\centering} m{0.03\textwidth}<{\centering} | m{0.3\textwidth}<{\centering}||m{0.03\textwidth}<{\centering} m{0.03\textwidth}<{\centering} | m{0.3\textwidth}<{\centering}|}
\hline
$a$ &$b$ & $S_{ab}$ & $a$ &$b$ & $S_{ab}$\\
\hline
$l$ & $l$ &$\displaystyle{\overset{\bar{l}}{[8]}[6]\overset{\bar{h}}{[2]} }$ & 
$L$ & $L$ &$\displaystyle{-[10]\overset{L}{[8]}[6]^2[4]\overset{H}{[2]} }$\\

$\bar{l}$ & $\bar{l}$ &$\displaystyle{\overset{l}{[8]}[6]\overset{h}{[2]} }$ &
$L$ & $h$ &$\displaystyle{\overset{l}{[10]}[8]^2 [6]^2 [4]^2 [2] }$\\

$l$ & $\bar{l}$ &$\displaystyle{-[10]\overset{L}{[6]}[4] }$ &
$L$ & $\bar{h}$ &$\displaystyle{\overset{\bar{l}}{[10]}[8]^2 [6]^2 [4]^2 [2] }$\\

$l$ & $L$ &$\displaystyle{\overset{l}{[9]}[7]\overset{h}{[5]}[3] }$ &
$L$ & $H$ &$\displaystyle{\overset{L}{[11]}[9]^2\overset{H}{[7]^3}[5]^3[3]^2[1]}$\\

$\bar{l}$ & $L$ &$\displaystyle{\overset{\bar{l}}{[9]}[7]\overset{\bar{h}}{[5]}[3] }$ &
$h$ & $h$ &$\displaystyle{\overset{\bar{l}}{[10]}\overset{\bar{h}}{[8]^3}[6]^3[4]^2[2]^2 }$\\

$l$ & $h$ &$\displaystyle{[9]\overset{\bar{h}}{[7]}[5]^2 [3]\overset{\bar{l}}{[11]} }$ &
$\bar{h}$ & $\bar{h}$ &$\displaystyle{\overset{l}{[10]}\overset{h}{[8]^3}[6]^3[4]^2[2]^2 }$\\

$\bar{l}$ & $\bar{h}$ &$\displaystyle{[9]\overset{h}{[7]}[5]^2 [3]\overset{l}{[11]} }$ &
$h$ & $\bar{h}$ &$\displaystyle{-[10]^2 [8]^2 \overset{H}{[6]^3}[4]^3[2] }$\\

$l$ & $\bar{h}$ &$\displaystyle{\overset{L}{[9]}[7]^2[5]\overset{H}{[3]}[1] }$ &
$h$ & $H$ &$\displaystyle{\overset{l}{[11]}\overset{h}{[9]^3}[7]^4[5]^4[3]^3[1] }$\\

$\bar{l}$ & $h$ &$\displaystyle{\overset{L}{[9]}[7]^2[5]\overset{H}{[3]}[1] }$ &
$\bar{h}$ & $H$ &$\displaystyle{\overset{\bar{l}}{[11]}\overset{\bar{h}}{[9]^3}[7]^4[5]^4[3]^3[1] }$\\

$l$ & $H$ &$\displaystyle{\overset{h}{[10]}[8]^2[6]^2[4]^2[2] }$ &
$H$ & $H$ &$\displaystyle{-\overset{L}{[10]^3}\overset{H}{[8]^5}[6]^6[4]^5[2]^3}$\\

$\bar{l}$ & $H$ &$\displaystyle{\overset{\bar{h}}{[10]}[8]^2[6]^2[4]^2[2] }$ & 
 & & \\
\hline
\end{tabular}
}
\caption{$S$-matrix of the $\epsilon$-TPM \cite{fateevCONFORMALFIELDTHEORY1990, sotkovBootstrapFusionsTricritical1989}. Here $[x]$ is a shorthand for $s_{x/12}(\theta) $
(see eq.(\ref{buildingblockk})). The superscript refers to the associated bound state. Notice that $12$ is the Coxeter number of the exceptional algebra $E_6$.}
\label{t_smatrix_TPM}
\end{table}
By looking at bound states with positive residue of two-particle scattering amplitudes it is possible to extract the mass spectrum of the model, which also coincides with the one of the $E_6$ Toda field theory eq.~\eqref{eq_masses_TPM}. 

In the $\epsilon$-TPM the degeneracy on the particle-antiparticle states $(l, \bar{l})$ and $(h, \bar{h})$ is lifted if we require that particle singlets and doublets transform according to the singlet and doublet irreducible representations of $S_3$ respectively.
In particular, we suppose that particles $l$ and $h$ have $Z_3$-charge $\omega$, $\bar{l}$ and $\bar{h}$ have $Z_3$-charge $\omega^*$, while $L$ and $H$ do not transform under the generator $\Omega$ of $S_3$. On the other hand, $Z_2$ charge conjugation maps particles to their respective antiparticles. In the specific case of the two singlets $L$ and $H$, particle and antiparticle coincide.

%%%%%%%%%%%%%

\section{Generalities on Form Factors} \label{s3}
The aim of this section is outlining the main properties of FFs, in particular the ones useful to address one and two-particle states. A further discussion of the analytic properties of such objects can be found in Appendix \ref{a1}.

\begin{figure}[t]
\centering
\tikzset{->-/.style={decoration={
  markings,
  mark=at position .5 with {\arrow{>}}},postaction={decorate}}}
\makebox[\textwidth][c]{
\begin{tikzpicture}
\node[draw, circle, minimum size = 36pt](O) at (0,1.5){$\mathcal{O}$};
\coordinate[ ](1) at (-2,-1){};
\coordinate[ ](2) at (-1,-1){};
\coordinate[ ](3) at (1,-1){};
\coordinate[ ](4) at (2,-1){};
\draw[->-] (1)node[below]{$\theta_1$}--(O.250);
\draw[->-] (2)node[below]{$\theta_2$}--(O.260);
\draw[->-] (3)node[below]{$\theta_{n-1}$}--(O.280);
\draw[->-] (4)node[below]{$\theta_n$}--(O.290);
\node[ ] at (0,0){...};
\end{tikzpicture}
}
\caption{Form factor of the operator $\mathcal{O}$.}
\label{f_ff_defi}
\end{figure}
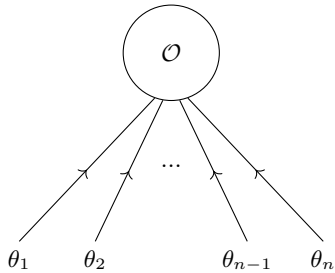

For a relativistic theory, a generic matrix element of an Hermitian operator $\mathcal{O}$ between an incoming and outgoing state is given by 
\begin{equation}
\label{eq_ff_translation}
\begin{split}
\prescript{}{OUT}{\langle} A_{{b}_1}(q_1)\dots &A_{{b}_m}(q_m)|\mathcal{O}(x)|A_{{a}_1}(p_1)\dots A_{{a}_n}(p_n)\rangle_{IN} =
\exp{i\left(\sum_{j=1}^m (q_j)_{\mu} - \sum_{i=1}^n (p_i)_{\mu}\right)x^\mu}\times\\
&\times \prescript{}{OUT}{\langle} A_{{b}_1}(q_1)\dots A_{{b}_m}(q_m)|\mathcal{O}(0)|A_{{a}_1}(p_1)\dots A_{{a}_n}(p_n)\rangle_{IN}.
\end{split}
\end{equation}
FFs are nothing more than matrix elements between the vacuum of the theory and incoming multi-particle states, as pictorially described in Fig.~\ref{f_ff_defi}. They are related to the previous equation by crossing, are analytic functions of the rapidity difference ${\theta_{ij} = \theta_i-\theta_j}$ and are defined as 
\begin{equation}\label{eq_ff_defi}
F_{{a}_1\dots {a}_n}^\mathcal{O}(\theta_1,\dots,\theta_n) = \langle 0|\mathcal{O}(0)|A_{{a}_1}(\theta_1)\dots A_{{a}_n}(\theta_n)\rangle.
\end{equation}
For a scalar operator, knowledge of FFs allows us to write the spectral series for the correlation functions which, for the two-point correlators, reads  \cite{zamolodchikovTwopointCorrelationFunction1991}
\begin{equation}
\label{eq_correlation_function_offcritical}
\langle\mathcal{O}(x)\mathcal{O}(0)\rangle = \sum_{n=0}^\infty \int \frac{\dd\theta_1\cdots\dd\theta_n}{(2\pi)^n n!}\abs{F_{a_1\dots a_n}^\mathcal{O}(\theta_1,\dots,\theta_n)}^2 \exp\left\{-\abs{x}\sum_{k=1}^n m_k \cosh\theta_k\right\}.
\end{equation}
An important feature of these spectral series is their rapid convergence \cite{Cardy:1993sx,mussardoStatisticalFieldTheory2020}. Practically, lowest mass particle FFs contribute more than higher mass particle ones, as it will be evidenced later by applications of the $\Delta$-theorem \cite{delfinoAsymptoticFactorisationForm1996} which is a sum rule for the conformal weight of an operator $\Phi$
\begin{equation}
\label{eq_delta_theorem}
\Delta_\Phi = -\frac{1}{4\pi \expval{\Phi}}\int\dd^2 x\ \expval{\Theta(x)\Phi(0)}.
\end{equation}
This sum rule is similar to Zamolodchikov's $c$-theorem sum rule for the UV central charge  \cite{zamolodchikovIrreversibilityFluxRenormalization1986} 
\begin{equation}
    \label{eq_ctheorem}
    c = \frac{3}{2}\int \dd r\ r^3\langle\Theta(r)\Theta(0)\rangle.
\end{equation}

For integrable theories, FFs of scalar operators can be parameterized as 
\begin{equation}
\label{eq_parametrization_ff}
F_{a_1\dots a_n}^\mathcal{O}(\theta_1,\dots,\theta_n) = \frac{Q_{a_1\dots a_n}^\mathcal{O}(\theta_1,\dots,\theta_n)}{D_{a_1\dots a_n}(\theta_1,\dots,\theta_n)}\prod_{i<j}F_{a_i a_j}^{min}(\theta_{ij}),
\end{equation}
where ${D_{a_1\dots a_n}(\theta_1,\dots,\theta_n)}$ is a $2\pi i$-periodic even function of the rapidity differences, ${Q_{a_1\dots a_n}^\mathcal{O}(\theta_1,\dots,\theta_n)}$ encodes the non-locality properties of $\mathcal{O}$ with respect to the operator generating particles, and $F_{a_i a_j}^{min}(\theta_{ij})$ are the so-called `minimal' FFs, which satisfy
\begin{equation}
\label{eq_minimal_ff_wilsoneq}
\begin{split}
&F^{min}_{ij}(\theta)= S_{ij}(\theta)F^{min}_{ji}(-\theta),\\
&F^{min}_{ij}(\theta+2\pi i) =F^{min}_{ji}(-\theta).
\end{split}
\end{equation}
Being relevant to Section \ref{s4}, the non-locality equations satisfied by FFs in the two-particle case read 
\begin{equation}
\label{eq_second_watson_2part}
F_{a_1 a_2}^{\mathcal{O}}(\theta) = S_{a_1 a_2}(\theta)e^{i2\pi \gamma_{a_2,\mathcal{O}}} F_{a_1 a_2}^{\mathcal{O}}(\theta+2\pi i),
\end{equation}
with $\theta = \theta_1 - \theta_2$.

FFs can be found by solving a set of linear equations in correspondence of kinematic and dynamical poles, for a suitable parametrization of ${Q_{a_1\dots a_n}^\mathcal{O}(\theta_1,\dots,\theta_n)}$. Concerning bound state poles, e.g. $|A_i(\theta) A_j (\theta+iu_{ij}^k)\rangle\to |A_k(iu_{ij}^k)\rangle$, a two-particle FF obeys 
\begin{equation}
\label{eq_simple_pole_2part}
-i\underset{\theta\to iu_{ij}^k}{\lim} (\theta-iu_{ij}^k)F^{\mathcal{O}}_{a_i,a_j}(\theta) = \Gamma_{ij}^k F^{\mathcal{O}}_{a_k},
\end{equation}
where $\Gamma_{ij}^k$ is the real three-point vertex 
\begin{equation}
\label{eq_three_point_vertex}
\Gamma_{ij}^k = \left[ -i \underset{\theta \to i u_{ij}^k}{\lim} (\theta - iu_{ij}^k) S_{ij} (\theta) \right]^{1/2}.
\end{equation}
There are also residue equations coming from double poles of the $S$-matrix. The equation for a pole located at ${i\phi}$ reads
\begin{equation}
\label{eq_double_pole_2part}
-i\underset{\theta\to i\varphi}{\lim}(\theta-i\varphi) F_{ij}(\theta) = \Gamma_{id}^c\Gamma_{\bar{d}j}^eF_{ce}(i\gamma),
\end{equation}
where $\gamma = \pi - u_{\bar{c}d}^{\bar{i}}-u_{d\bar{e}}^{\bar{j}}$ ($\bar{j}$ denoting the antiparticle of $j$). The interested reader can find a thorugh discussion of these equations in Refs. \cite{acerbiFormFactorsCorrelation1996, delfinoSpinspinCorrelationFunction1995}.
Kinematic poles are due to particle-antiparticle annihilation and in these cases two-particle FFs obey \cite{yurovCORRELATIONFUNCTIONSINTEGRABLE1991}
\begin{equation}
\label{eq_kinetic_residue_2part}
-i\underset{\theta\to i\pi}{\lim}(\theta-i\pi)F^{\mathcal{O}}_{a_i,a_{\bar{i}}}(\theta) = \left[1-e^{i2\pi \gamma_{a_i,\mathcal{O}}}\right]\langle \mathcal{O}\rangle,
\end{equation} 
where $\langle \mathcal{O}\rangle$ is the vacuum expectation value (VEV) of the operator $\mathcal{O}$.

In the two-particle case, ${Q_{a_i a_j}^\mathcal{O}(\theta)}$ is a polynomial in $e^\theta$, which satisfies an upper bound on its maximal degree \cite{delfinoSpinspinCorrelationFunction1995}. Define $y_{\mathcal{O}}$ to be the coefficient multiplying $\theta_i$ as it approaches infinity, i.e.
\begin{equation}
F_{a_i a_j}^\mathcal{O}(\theta_i,\theta_j)\underset{|\theta_i|\to \infty}{\sim}  \exp\left\{y_{\mathcal{O}}|\theta_{i}|\right\}.
\end{equation}
The upper bound on the degree of the FF reads
\begin{equation}
\label{eq_bound_poly_order}
y_{\mathcal{O}}\leq \Delta_{\mathcal{O}},
\end{equation}
where $\Delta_{\mathcal{O}}$ is the conformal weight of the operator $\mathcal{O}$ in the UV CFT. 

The latter bound and the aforementioned linear equations are in most cases sufficient to fix the FFs as functions of the VEV of some operator. However, it may happen that they are not enough. In computing the FFs of order and disorder, leading and subleading, operators of the thermally deformed TPM, we also made use of second order equations coming from the asymptotic factorization of FFs \cite{delfinoAsymptoticFactorisationForm1996}.

%%%%%%%%%%%%%%%%%%%%%

\section{Form Factors of the Stress-Energy Tensor}\label{s5}

This section outlines the computation of the FFs of the stress-energy tensor $\Theta(x)$, as originally obtained in \cite{acerbiFormFactorsCorrelation1996}.

Since the FFs of the stress-energy tensor corresponding to the same particle states have all the same pole structure, the two-particle FFs of the trace $\Theta$ can be parametrized by the following polynomial:
\begin{equation}
    \label{eq_Q_poly_stressenergy}
    Q^{\Theta}_{ab}(\theta) = \left(\cosh\theta+\frac{m_a^2+m_b^2}{2m_a m_b}\right)^{1-\delta_{m_a,m_b}}\sum_{i=0}^{\deg Q^\Theta_{ab}}c_{ab}^{i}(\cosh\theta)^i,
\end{equation}
where the degree is determined by the asymptotic behavior eq.~\eqref{eq_bound_poly_order}. Notice that when particles have the same mass, i.e. in the case of a particle-antiparticle doublet or a singlet, the prefactor in the latter equation is completely absorbed in the unknown coefficients $c_{ab}^{(i)}$. Being conserved quantities of the theory, the FFs of $\Theta$ do not obey the usual residue equations for particle-antiparticle annihilation, rather:
\begin{equation}
\label{eq_stressenergy_normalization}
    F_{a\bar{a}}^{\Theta}(i\pi) = 2\pi m_a^2.
\end{equation}

Apart from these properties, $\Theta$ is a $Z_3$-invariant operator, like $\mu$ and $\zeta$. The problem of finding the one and two-particle FFs is therefore similar for those three operators. In the following we'll present the parametrization and the equations satisfied by the one-particle and first two-particle FFs (Table \ref{t_disorder_lowestmass}).
\begin{table}
\centering
\def\arraystretch{1.5}
\begin{tabular}{| c | l |}
\hline
state & $E_{CM}/m_l$\\
\hline
\hline
$L$ & $1.41421\dots$\\
\hline
$l\bar{l},\bar{l}l$ & $\geq 2$\\
\hline
$H$ & $ 2.73205 \dots $\\
\hline
$LL$ & $\geq 2.82843 \dots$\\
\hline
$l\bar{h},\bar{l}h,\bar{h}l, h\bar{l}$ & $\geq 2.93185 \dots$\\
\hline
$lll, \bar{l}\bar{l}\bar{l}$ & $\geq 3$\\
\hline
\end{tabular}
\caption{Multi-particle states with lowest center of mass energy and nontrivial FFs with respect to the $Z_3$-invariant operators.}
\label{t_disorder_lowestmass}
\end{table}
The first two-particle FFs of $\Theta$ are parametrized as follows
\begin{equation}
\label{eq_stressenergy_FFs}
\begin{split}
F_{l\bar{l}}^\Theta (\theta) &=c_{l\bar{l}}^0\frac{F_{l\bar{l}}^{min}(\theta)}{D_{l\bar{l}}(\theta)},\\
F_{LL}^\Theta (\theta) &=\left(c_{LL}^0+c_{LL}^1\cosh\theta\right)\frac{F_{LL}^{min}(\theta)}{D_{LL}(\theta)},\\
F_{l\bar{h}}^\Theta (\theta) &=c_{l\bar{h}}^0\left(\cosh\theta + \frac{m_l^2 + m_h^2}{2 m_l m_h}\right)\frac{F_{l\bar{h}}^{min}(\theta)}{D_{l\bar{h}}(\theta)},
\end{split}
\end{equation}
and satisfy the equations
\begin{align}
&F_{l\bar{l}}^{\Theta}(i\pi) = 2\pi m_l^2,\\
&-i\underset{\theta\to iu_{l\bar{l}}^L}{\lim} (\theta-iu_{l\bar{l}}^L)F_{l\bar{l}}^{\Theta}(\theta) = \Gamma_{l\bar{l}}^L F_L^\Theta,\\
&F_{LL}^{\Theta}(i\pi) = 2\pi m_L^2,\\
&-i\underset{\theta\to iu_{LL}^L}{\lim} (\theta-iu_{LL}^L)F_{LL}^{\Theta}(\theta) = \Gamma_{LL}^L F_L^\Theta,\\
&-i\underset{\theta\to iu_{LL}^H}{\lim} (\theta-iu_{LL}^H)F_{LL}^{\Theta}(\theta) = \Gamma_{LL}^H F_H^\Theta,\\
&-i\underset{\theta\to i\pi /2}{\lim} (\theta-i\pi /2)F_{LL}^{\Theta}(\theta) = \Gamma_{Ll}^{l}\Gamma_{L\bar{l}}^{\bar{l}} F_{l\bar{l}}^\Theta (0),\\
&-i\underset{\theta\to iu_{l\bar{h}}^L}{\lim} (\theta-iu_{l\bar{h}}^L)F_{l\bar{h}}^{\Theta}(\theta) = \Gamma_{l\bar{h}}^L F_L^\Theta,\\
&-i\underset{\theta\to iu_{l\bar{h}}^H}{\lim} (\theta-iu_{l\bar{h}}^H)F_{l\bar{h}}^{\Theta}(\theta) = \Gamma_{l\bar{h}}^H F_H^\Theta,\\
&-i\underset{\theta\to i\pi 7/12}{\lim} (\theta-i\pi 7/12)F_{l\bar{h}}^{\Theta}(\theta) = \Gamma_{ll}^{\bar{l}} \Gamma_{\bar{h}\bar{l}}^{l}F_{\bar{l}l}^\Theta (i\pi 2/12).
\end{align}
Solutions of this system of equations are provided in Table \ref{t_stressenergy_summary}, and reproduce the ones of \cite{acerbiFormFactorsCorrelation1996}. As a further check of consistency, we computed the contributions to the $c$-theorem eq.~\eqref{eq_ctheorem} (Table \ref{t_contrib_c}).
\begin{table}
\centering
\def\arraystretch{1.5}
\begin{minipage}{.45\textwidth}
\begin{tabular}{| c | l |}
\hline
Param.$/m_l^2$ & \\
\hline
$F_{L}^\Theta$ &
$1.2613539473\dots$\\
\hline
$c_{l\bar{l}}^0$ & $6.2831853071\dots$\\
\hline
$F_H^\Theta$ & $0.2920374047\dots$\\
\hline
$c_{LL}^0$ & $21.765592370\dots$\\
$c_{LL}^1$ & $9.1992217564\dots$\\
\hline
$c_{l\bar{h}}^0$ & $25.226482640\dots$\\
\hline
\end{tabular}
\end{minipage}
\begin{minipage}{.45\textwidth}
\begin{tabular}{| c | l |}
\hline
Param./$\langle\epsilon\rangle$ & \\
\hline
$F_{L}^\epsilon$ &
$-0.9499625864\dots$\\
\hline
$d_{l\bar{l}}^0$ & 
$-4.7320508076\dots$\\
\hline
$F_H^\epsilon$ & 
$-0.2199419194\dots$\\
\hline
$d_{LL}^0$ & 
$-16.392304845\dots$\\
$d_{LL}^1$ & 
$-6.9282032303\dots$\\
\hline
$d_{l\bar{h}}^0$ & 
$-18.998802632\dots$\\
\hline
\end{tabular}
\end{minipage}
\caption{Parameters of the first FFs of the trace of the stress-energy tensor $\Theta$ and the energy operator $\epsilon$.}
\label{t_stressenergy_summary}
\end{table}
\begin{table}[t]
\centering
\def\arraystretch{1.5}
\begin{tabular}{| c | l |}
\hline
State & $c$-series\\
\hline
\hline
$A_L$           & $0.7596528$\\
\hline
$A_{l\bar{l}}$  & $0.0853913$\\
\hline
$A_H$           & $0.0029236$\\
\hline
$A_{LL}$        & $0.0024957$\\
\hline
$A_{l\bar{h}}$  & $0.0049010$\\
\hline
\hline
Total & 0.85536...\\
\hline
\hline
Relative error & $0.207\%$\\
\hline
\end{tabular}
\caption{$c$-theorem contributions of the first five particle states ($c = 6/7 \approx 0.85714$).}
\label{t_contrib_c}
\end{table}

Since the fundamental mass $m_l$ obeys eq.~\eqref{eq_stressenergy_normalization}, it is possible to compute $\langle \Theta \rangle$ from the cluster equation \cite{Koubek:1993ke}
\begin{equation}
\label{eq_clusterLL_stressenergy}
\underset{\theta\to\infty}{\lim} F_{LL}^\Theta (\theta) = \frac{w_{LL}}{\langle \Theta\rangle} F_{L}^\Theta F_L^\Theta,
\end{equation}
where $w_{LL}$ is a phase, which can be either $1$ or $-1$. In order to make the $\Delta$-theorem contributions for the energy operator $\epsilon$ positive, we choose the phase $w_{LL}$ such that $\langle \Theta\rangle$ is negative, yielding
\begin{equation}
\label{eq_vev_stressenergy_conformalnormalization}
    \langle \Theta\rangle = -1.3277932893\dots m_l^2.
\end{equation}
Combining Eqs.~\eqref{eq_trace_stress_deformation} and \eqref{eq_mass_lightestkink} we get
\begin{equation}
\label{eq_vev_stressenergy_conformalnormalization}
    \langle \epsilon \rangle = -1.7303450451\dots \lambda^{1/6},
\end{equation}
which coincides with the VEV of the energy operator in the thermal perturbation of the whole minimal model $\mathcal{M}_{6,7}$ in the conformal normalization \cite{fateevExpectationValuesLocal1998}.

The FFs of $\epsilon$ retain the same pole structure and polynomial dependence of the ones of $\Theta$, with modified coefficients; in particular, the FFs are parametrized as follows
\begin{equation}
\label{eq_energy_FFs}
\begin{split}
F_{l\bar{l}}^\epsilon (\theta) &=d_{l\bar{l}}^0\frac{F_{l\bar{l}}^{min}(\theta)}{D_{l\bar{l}}(\theta)},\\
F_{LL}^\epsilon (\theta) &=\left(d_{LL}^0+d_{LL}^1\cosh\theta\right)\frac{F_{LL}^{min}(\theta)}{D_{LL}(\theta)},\\
F_{l\bar{h}}^\epsilon (\theta) &=d_{l\bar{h}}^0\left(\cosh\theta + \frac{m_l^2 + m_h^2}{2 m_l m_h}\right)\frac{F_{l\bar{h}}^{min}(\theta)}{D_{l\bar{h}}(\theta)},
\end{split}
\end{equation}
with the unknown coefficients satisfying
\begin{equation}
    d_{ab}^i =  \frac{c_{ab}^i}{\langle\Theta\rangle}\langle\epsilon\rangle.
\end{equation}
For the sake of completeness, their values are summarized in Table \ref{t_stressenergy_summary} and the contributions of the respective FFs in the $\Delta$ sum rule are shown in Table \ref{t_contrib_delta_energy}.
\begin{table}
\centering
\def\arraystretch{1.5}
\begin{tabular}{| c | l |}
\hline
State & $\Delta_\epsilon$-series\\
\hline
\hline
$A_L$           & $0.0953527$\\
\hline
$A_{l\bar{l}}$  & $0.0313356$\\
\hline
$A_H$           & $0.0013696$\\
\hline
$A_{LL}$        & $0.0026516$\\
\hline
$A_{l\bar{h}}$  & $0.0046187$\\
\hline
\hline
Total & 0.135328 \\
\hline
\hline
Relative error & $5.27\%$\\
\hline
\end{tabular}
\caption{$\Delta$-theorem contributions for the energy operator $\epsilon$ of the first five particle states ($\Delta_\epsilon = 1/7 \approx 0.14286$).}
\label{t_contrib_delta_energy}
\end{table}

%%%%%%%%%%%%%%%%%%%%%%

\section{Form Factors of Order and Disorder Leading and Subleading Operators} \label{s4}

This section is dedicated to the derivation of the first FFs of magnetic operators in the thermally deformed TPM. Let us start from the leading disorder operator.

\subsection{Leading Disorder Operator $\mu$}\label{s4_1}

Because of $\mathcal{C}$-invariance, $\langle \tilde{\mu} \rangle $ and $\langle \mu \rangle$ coincide. In Table \ref{t_disorder_lowestmass} are presented the first states with lowest mass and nontrivial FFs. Since the computation of these FFs involves non-locality factors, the derivation will be outlined step-by-step.

Let us start from ${F_{l\bar{l}}^\mu(\theta)}$. Since in the end we are concerned with computing the off-critical two-point correlation functions of the operators, it is important to notice that the only contribution that needs to be evaluated numerically in eq.~\eqref{eq_correlation_function_offcritical} is that of ${F_{l\bar{l}}^\mu(\theta)}$. Indeed, because of $\mathcal{C}$-invariance, the FFs of the operator $\tilde{\mu}$ are read from that of $\mu$:
\begin{equation}
F_{\bar{l} l}^{\tilde{\mu}}(\theta) = F_{l\bar{l}}^{\mu}(\theta), \quad F_{l \bar{l}}^{\tilde{\mu}}(\theta) = F_{\bar{l} l}^{\mu}(\theta) = S_{l\bar{l}}(\theta) F_{l\bar{l}}^{\mu}(-\theta),
\end{equation}
where the last equality is a consequence of eq.~\eqref{eq_minimal_ff_wilsoneq}.

Since the disorder operator has nonzero non-locality factor with respect to particles $l$ and $\bar{l}$, the FF will also possess an additional term which takes into account particle-antiparticle annihilation. Because of the asymptotic behaviors of ${F_{l\bar{l}}^{min}(\theta)}$ and ${D_{l\bar{l}}(\theta)}$, $Q_{l\bar{l}}^\mu(\theta)$ satisfies the following inequality on its degree: 
\begin{equation}
\label{eq_lbl_degree_inequality}
(\deg Q_{l\bar{l}}^\mu)\theta-\frac{5}{6}|\theta|\leq \frac{1}{21}|\theta|.
\end{equation}
Notice the absence of the absolute value on the first term in the l.h.s.. The polynomial ${Q_{l\bar{l}}^\mu(\theta)}$ is not symmetric in the exchange ${\theta\to -\theta}$, since the non-locality factors of $\mu$ with respect to particles $l$ and $\bar{l}$ are nonzero. The non-locality equation that must be satisfied is
\begin{equation}
F_{l\bar{l}}^{\mu}(\theta) = S_{l\bar{l}}(\theta)e^{i2\pi \gamma_{\bar{l},\mu}} F_{l\bar{l}}^{\mu}(\theta+2\pi i),
\end{equation} 
with ${\gamma_{\bar{l},\mu}=-1/3}$.
Solutions are sought of the form $\exp(\alpha\theta)$, provided that they satisfy
\begin{equation}
e^{\alpha\theta} = -e^{-i2\pi/3}e^{\alpha(\theta+2\pi i)}.
\end{equation}
Therefore the parametrization for this FFs is 
\begin{equation}
\label{eq_FlBl}
F_{l\bar{l}}^\mu(\theta) =\left(a_{l\bar{l}}^0\ e^{\frac{5}{6}\theta}+a_{l\bar{l}}^1\ e^{-\frac{1}{6}\theta}\right)\frac{F_{l\bar{l}}^{min}(\theta)}{\cosh\left(\frac{\theta}{2}\right) D_{l\bar{l}}(\theta)}.
\end{equation}

Pole equations 
\begin{equation}
\label{eq_pole_lBl}
\begin{split}
&-i\underset{\theta\to i\pi}{\lim} (\theta-i\pi)F_{l\bar{l}}^{\mu}(\theta) = \left[-i\underset{\theta\to i\pi}{\lim} (\theta-i\pi)F_{\bar{l}l}^{\mu}(\theta)\right]^*= \left(1-e^{i2\pi/3}\right)\langle \mu\rangle,\\
&-i\underset{\theta\to iu_{l\bar{l}}^L}{\lim} (\theta-iu_{l\bar{l}}^L)F_{l\bar{l}}^{\mu}(\theta) = -i\underset{\theta\to iu_{l\bar{l}}^L}{\lim} (\theta-iu_{l\bar{l}}^L)F_{\bar{l}l}^{\mu}(\theta) = \Gamma_{l\bar{l}}^L F_L^\mu,
\end{split}
\end{equation}
allow us fixing all three unknowns - $a_{l\bar{l}}^0$, $a_{l\bar{l}}^1$ and $F_L^\mu$ - as functions of the VEV of the leading disorder operator $\mu$ (Table \ref{t_leading_disorder_summary}).

As far as ${F_{LL}^\mu(\theta)}$ is concerned, the absence of non-locality phases simplifies the parametrization of the FF:
\begin{equation}
\label{eq_FLL}
F_{LL}^\mu(\theta) =\left(a_{LL}^0+a_{LL}^1\cosh\theta\right)\frac{F_{LL}^{min}(\theta)}{D_{LL}(\theta)}.
\end{equation}
However, the corresponding $S$-matrix has a second-order pole; therefore the set of pole equations is 
\begin{equation}
\label{eq_pole_LL}
\begin{split}
&-i\underset{\theta\to iu_{LL}^L}{\lim} (\theta-iu_{LL}^L)F_{LL}^{\mu}(\theta) = \Gamma_{LL}^L F_L^\mu,\\
&-i\underset{\theta\to iu_{LL}^H}{\lim} (\theta-iu_{LL}^H)F_{LL}^{\mu}(\theta) = \Gamma_{LL}^H F_H^\mu,\\
&-i\underset{\theta\to i\pi /2}{\lim} (\theta-i\pi /2)F_{LL}^{\mu}(\theta) = \Gamma_{L\bar{l}}^{\bar{l}} \Gamma_{Ll}^{l}F_{\bar{l}l}^\mu (0) = \Gamma_{Ll}^{l}\Gamma_{L\bar{l}}^{\bar{l}} F_{l\bar{l}}^\mu (0).
\end{split}
\end{equation}
Again, all three unknowns ($F_H^\mu$, $a_{LL}^0$ and $a_{LL}^1$) can be fixed by simply solving these linear equations (Table \ref{t_leading_disorder_summary}). The first nontrivial check of the consistency of these solutions is given by the second order cluster equation
\begin{equation}
\label{eq_clusterLL}
\underset{\theta\to\infty}{\lim} F_{LL}^\mu (\theta) = \frac{w_{LL}}{\langle \mu\rangle} F_{L}^\mu F_L^\mu,
\end{equation}
satisfied when the phase $w_{LL} = 1$.

The FF $F_{l\bar{h}}^\mu(\theta)$ is obtained along the lines of ${F_{l\bar{l}}^\mu(\theta)}$. In particular, its parametrization reads 
\begin{equation}
\label{eq_FlBh}
F_{l\bar{h}}^\mu(\theta) =\left(a_{l\bar{h}}^0\ e^{\frac{4}{3}\theta}+a_{l\bar{h}}^1\ e^{\frac{1}{3}\theta}+a_{l\bar{h}}^2\ e^{-\frac{2}{3}\theta}\right)\frac{F_{l\bar{h}}^{min}(\theta)}{D_{l\bar{h}}(\theta)},
\end{equation}
and satisfies the pole equations \begin{equation}
\label{eq_pole_lBh}
\begin{split}
&-i\underset{\theta\to iu_{l\bar{h}}^L}{\lim} (\theta-iu_{l\bar{h}}^L)F_{l\bar{h}}^{\mu}(\theta) = \Gamma_{l\bar{h}}^L F_L^\mu,\\
&-i\underset{\theta\to iu_{l\bar{h}}^H}{\lim} (\theta-iu_{l\bar{h}}^H)F_{\bar{h}l}^{\mu}(\theta) = \Gamma_{l\bar{h}}^H F_H^\mu,\\
&-i\underset{\theta\to i\pi 7/12}{\lim} (\theta-i\pi 7/12)F_{l\bar{h}}^{\mu}(\theta) = \Gamma_{ll}^{\bar{l}} \Gamma_{\bar{l}\bar{l}}^{l}F_{\bar{l}l}^\mu (i\pi 2/12),\\
&-i\underset{\theta\to i\pi 7/12}{\lim} (\theta-i\pi 7/12)F_{\bar{h}l}^{\mu}(\theta) = \Gamma_{ll}^{\bar{l}} \Gamma_{\bar{l}\bar{l}}^{l}F_{l\bar{l}}^\mu (i\pi 2/12).
\end{split}
\end{equation}
Notice that the latter two pole equations have a peculiar dependence on the two-particle FFs ${F_{\bar{l}l}^\mu (\theta)}$ and ${F_{l\bar{l}}^\mu (\theta)}$. Solutions are reported in Table \ref{t_leading_disorder_summary}. We also computed the first contributions to the $\Delta$-theorem (Table \ref{t_contrib_delta_leading}).
\begin{table}
\centering
\def\arraystretch{1.5}
\begin{minipage}{0.45\textwidth}
\begin{tabular}{| c | l |}
\hline
Param.$/\langle\mu\rangle$ & \\
\hline
$F_L^\mu$ & $-0.3477104392\dots$\\
\hline
$a_{l\bar{l}}^0$ & $-0.3169872981\dots$\\
$a_{l\bar{l}}^1$ & $-1.1830127018\dots$\\
\hline
$F_H^\mu$ & $-0.0589332596\dots$\\
\hline
$a_{LL}^0$ & $-5.1961524220\dots$\\
$a_{LL}^1$ & $-0.9282032290\dots$\\
\hline
$a_{l\bar{h}}^0$ & $ -0.6587868452\dots$\\
$a_{l\bar{h}}^1$ & $-6.0223791136\dots$\\
$a_{l\bar{h}}^2$ & $-4.2584651101\dots$\\
\hline
\end{tabular}
\end{minipage}
\begin{minipage}{0.45\textwidth}
\begin{tabular}{| c | l |}
\hline
Param.$/\langle\zeta\rangle$ & \\
\hline
$F_{L}^\zeta$ & $-2.2476356112\dots$\\
\hline
$b_{l\bar{l}}^0$ & $-4.4150635077\dots$\\
$b_{l\bar{l}}^1$ & $-7.6471143143\dots$\\
$b_{l\bar{l}}^2$ & $-2.3660254027\dots$\\
\hline
$F_H^\zeta$ & $-0.8408344177\dots$\\
\hline
$b_{LL}^0$ & $-49.9807620886\dots$\\
$b_{LL}^1$ & $-38.7646096627\dots$\\
\hline
$b_{l\bar{h}}^0$ & $-22.6098992318\dots$\\
$b_{l\bar{h}}^1$ & $-74.3815188798\dots$\\
$b_{l\bar{h}}^2$ & $-59.3127674978\dots$\\
$b_{l\bar{h}}^3$ & $-9.4994013120\dots$\\
\hline
\end{tabular}
\end{minipage}
\caption{Parameters of the first FFs of the leading disorder operator $\mu$ and subleading disorder operator $\zeta$.}
\label{t_leading_disorder_summary}
\end{table}
\begin{table}
\centering
\def\arraystretch{1.5}
\begin{tabular}{| c | l | l|}
\hline
State & $\Delta_{\mu}$-series & $\Delta_\zeta$-series\\
\hline
\hline
$A_L$           & $0.0349015$ & $0.2256069$\\
\hline
$A_{l\bar{l}}$  & $0.0095534$  & $0.1266482$\\
\hline
$A_H$           & $0.0003670$  & $0.0051113$\\
\hline
$A_{LL}$        & $0.0005463$ & $0.0121824$\\
\hline
$A_{l\bar{h}}$  & $0.0010622$ & $0.0257981$\\
\hline
\hline
Total & 0.0464304 & 0.395347\\
\hline
\hline
Relative error & $2.50\%$ & $16.98\%$\\
\hline
\end{tabular}
\caption{$\Delta$-theorem contributions for the leading disorder operator $\mu$ ($\Delta_\mu = 1/21 \approx 0.047619$) and the subleading disorder operator $\zeta$ of the first five particle states ($\Delta_\zeta = 10/21 \approx 0.47619$).}
\label{t_contrib_delta_leading}
\end{table}

\subsection{Subleading Disorder Operator $\zeta$}
The computation of FFs for the subleading disorder operator $\zeta$ presents little differences if compared to the case of $\mu$. First of all, since the $\zeta$ has the same transformation properties as $\mu$, the non-vanishing FFs of $\zeta$ are the same as the ones of $\mu$ (Table \ref{t_disorder_lowestmass}). The only difference in computing the FFs between the leading and subleading disorder operator lies in their different anomalous dimensions, which in turn affects the upper bound on the FFs degree. On the other hand non-locality coefficients remain unchanged in passing from the leading to the subleading case. These considerations are accounted for in the parametrization of some of the $Q$ polynomials:
\begin{equation}
\label{eq_disorder_subleading_FFs}
\begin{split}
F_{l\bar{l}}^\zeta (\theta) &=\left(b_{l\bar{l}}^0\ e^{\frac{5}{6}\theta}+b_{l\bar{l}}^1\ e^{-\frac{1}{6}\theta}+b_{l\bar{l}}^2\ e^{-\frac{7}{6}\theta}\right)\frac{F_{l\bar{l}}^{min}(\theta)}{\cosh\left(\frac{\theta}{2}\right) D_{l\bar{l}}(\theta)},\\
F_{LL}^\zeta (\theta) &=\left(b_{LL}^0+b_{LL}^1\cosh\theta\right)\frac{F_{LL}^{min}(\theta)}{D_{LL}(\theta)},\\
F_{l\bar{h}}^\zeta (\theta) &=\left(b_{l\bar{h}}^0\ e^{\frac{4}{3}\theta}+b_{l\bar{h}}^1\ e^{\frac{1}{3}\theta}+b_{l\bar{h}}^2\ e^{-\frac{2}{3}\theta}+b_{l\bar{h}}^3\ e^{-\frac{5}{3}\theta}\right)\frac{F_{l\bar{h}}^{min}(\theta)}{D_{l\bar{h}}(\theta)}.
\end{split}
\end{equation}
The set of linear equations for $\zeta$ remains the same as the ones for $\mu$, except for a change in label $\mu\to \zeta$ in all of them. However, it turns out that there is one linearly dependent equation, which doesn't allow fixing one of the unknowns, e.g. $F_{L}^\zeta$. Studying the cluster equation \cite{Koubek:1993ke}
\begin{equation}
\label{eq_clusterLL_subleading}
\underset{\theta\to\infty}{\lim} F_{LL}^\zeta (\theta) = \frac{w_{LL}}{\langle \zeta\rangle} F_{L}^\zeta F_L^\zeta,
\end{equation}
provides a solution to the problem. If ${w_{LL} = 1}$, we have two nonphysical solutions
\begin{equation*}
F_{L}^\zeta = (-0.2725122015\dots, 2.8678582521\dots)\langle \zeta\rangle,
\end{equation*}
since the first value gives a contribution to the $\Delta$-theorem which is positive but too small, while the second gives a negative contribution. If ${w_{LL} = -1}$, one finds
\begin{equation*}
F_{L}^\zeta = (-2.2476356112\dots, -0.3477104393\dots)\langle \zeta\rangle.
\end{equation*}
While the second solution is associated to the disorder operator $\mu$, the first one is acceptable: its contribution to the $\Delta$-theorem is given in Table \ref{t_contrib_delta_leading} and it is compatible with the conformal weight of the operator. Notice that convergence is slower than the leading case, as it generally happens when one considers less relevant operators - see also Ref. \cite{cortescuberoDualityFormFactors2022}. Solving the system of equations, together with eq.~\eqref{eq_clusterLL_subleading}, yields the results of Table \ref{t_leading_disorder_summary}.

\subsection{Leading Order Operator $\sigma$ and Subleading Order Operator $Z$}

The FFs of the leading and subleading order operators are simpler to compute if compared to their disorder counterparts, because there are no non-locality factors involved. The first nontrivial FFs are summarized in Table \ref{t_order_lowestmass}. It turns out that the solution of their pole equations require the knowledge of the FFs of disorder operators.
\begin{table}
\centering
\def\arraystretch{1.5}
\begin{tabular}{| c | l |}
\hline
state & $E_{CM}/m_l$\\
\hline
\hline
$\bar{l}$ & $1$\\
\hline
$\bar{h}$ & $1.93185\dots$\\
\hline
$ll$ & $\geq 2$\\
\hline
$\bar{l}L$ & $\geq 2.41421 \dots $\\
\hline
$lh$ & $\geq 2.93185 \dots$\\
\hline
$\bar{l}l\bar{l}$ & $\geq 3$\\
\hline
$\bar{h}L$ & $\geq 3.34606 \dots$\\
\hline
\end{tabular}
\caption{Multi-particle states with lowest center of mass energy and nontrivial FF with respect to the order operator $\sigma$. The ones of the subleading order operator $Z$ are retrieved by charge-conjugating all particles.}
\label{t_order_lowestmass}
\end{table}

The parametrizations of the first FFs of $\sigma$ are
\begin{equation}
\label{eq_order_leading_FFs}
\begin{split}
F_{ll}^\sigma(\theta) &=a_{ll}^0 \frac{F_{ll}^{min}(\theta)}{D_{ll}(\theta)},
\\
    F_{\bar{l}L}^\sigma(\theta) &=\left(a_{\bar{l}L}^0+a_{\bar{l}L}^1 \cosh\theta \right) \frac{F_{\bar{l}L}^{min}(\theta)}{D_{\bar{l}L}(\theta)},\\
    F_{lh}^\sigma(\theta) &=\left(a_{lh}^0+a_{lh}^1 \cosh\theta \right) \frac{F_{lh}^{min}(\theta)}{D_{lh}(\theta)},\\
    F_{\bar{h}L}^\sigma(\theta) &=\left(a_{\bar{h}L}^0+a_{\bar{h}L}^1 \cosh\theta+ a_{\bar{h}L}^2 (\cosh\theta)^2 \right) \frac{F_{\bar{h}L}^{min}(\theta)}{D_{\bar{h}L}(\theta)}.
\end{split}
\end{equation}
and the set of pole equations is 
\begin{align}
&-i\underset{\theta\to iu_{ll}^{\bar{l}}}{\lim} (\theta-iu_{ll}^{\bar{l}})F_{ll}^{\sigma}(\theta) = \Gamma_{ll}^{\bar{l}} F_{\bar{l}}^\sigma,\\
&-i\underset{\theta\to iu_{ll}^{\bar{h}}}{\lim} (\theta-iu_{ll}^{\bar{h}})F_{ll}^{\sigma}(\theta) = \Gamma_{ll}^{\bar{h}} F_{\bar{h}}^\sigma,\\
    &-i\underset{\theta\to iu_{\bar{l}L}^{\bar{l}}}{\lim} (\theta-iu_{\bar{l}L}^{\bar{l}})F_{\bar{l}L}^{\sigma}(\theta) = \Gamma_{\bar{l}L}^{\bar{l}} F_{\bar{l}}^\sigma,\\
&-i\underset{\theta\to iu_{\bar{l}L}^{\bar{h}}}{\lim} (\theta-iu_{\bar{l}L}^{\bar{h}})F_{\bar{l}L}^{\sigma}(\theta) = \Gamma_{\bar{l}L}^{\bar{h}} F_{\bar{h}}^\sigma,\\
&-i\underset{\theta\to iu_{lh}^{\bar{l}}}{\lim} (\theta-iu_{lh}^{\bar{l}})F_{lh}^{\sigma}(\theta) = \Gamma_{lh}^{\bar{l}} F_{\bar{l}}^\sigma,\\
&-i\underset{\theta\to iu_{lh}^{\bar{h}}}{\lim} (\theta-iu_{lh}^{\bar{h}})F_{lh}^{\sigma}(\theta) = \Gamma_{lh}^{\bar{h}} F_{\bar{h}}^\sigma,\\
&-i\underset{\theta\to i\pi 5/12}{\lim} (\theta-i\pi 5/12)F_{lh}^{\sigma}(\theta) = \Gamma_{lh}^{\bar{l}}\Gamma_{l\bar{l}}^{L}F_{\bar{l}L}^{\sigma}(i\pi/12),\\
&-i\underset{\theta\to iu_{\bar{h}L}^{\bar{l}}}{\lim} (\theta-iu_{lh}^{\bar{l}})F_{\bar{h}L}^{\sigma}(\theta) = \Gamma_{\bar{h}L}^{\bar{l}} F_{\bar{l}}^\sigma,\\
&-i\underset{\theta\to i\pi 8/12}{\lim} (\theta-i\pi 8/12)F_{\bar{h}L}^{\sigma}(\theta) = \Gamma_{\bar{l}\bar{h}}^{l}\Gamma_{lL}^{l}F_{ll}^{\sigma}(i\pi 4/12),\\
&-i\underset{\theta\to i\pi 4/12}{\lim} (\theta-i\pi 4/12)F_{\bar{h}L}^{\sigma}(\theta) = \Gamma_{\bar{l}\bar{h}}^{l}\Gamma_{lL}^{h}F_{lh}^{\sigma}(i\pi/12),\\
&-i\underset{\theta\to i\pi 6/12}{\lim} (\theta-i\pi 6/12)F_{\bar{h}L}^{\sigma}(\theta) = \Gamma_{\bar{h}L}^{\bar{l}}\Gamma_{l\bar{l}}^{L}F_{\bar{l}L}^{\sigma}(i\pi/12).
\end{align}
These equations are linearly dependent and $F_{\bar{l}}^\sigma$ cannot be fixed in this way. This value, however, can be obtained employing
 the cluster equation 
\begin{equation}
\label{eq_clusterlBl}
\underset{\theta\to+\infty}{\lim} F_{l\bar{l}}^\mu(\theta) = \frac{1}{\langle \mu\rangle}F_l^{\bar{\sigma}}F_{\bar{l}}^\sigma = \frac{1}{\langle \mu\rangle}\left(F_{\bar{l}}^\sigma\right)^2
\,\,\,.
\end{equation}
This equation allows us to fix $F_{\bar{l}}^\sigma$ up to a sign, which however is not relevant to the computation of two-point correlation functions. 
The parameters are summarized in Table \ref{t_leading_order_summary}.
\begin{table}
\centering
\def\arraystretch{1.5}
\begin{minipage}{0.45\textwidth}
\begin{tabular}{| c | l |}
\hline
Param.$/\langle\mu\rangle$ & \\
\hline
$F_{\bar{l}}^\sigma$ & $0.6143373207\dots$\\
$F_{\bar{h}}^\sigma$ & $-0.1788394265\dots$\\
\hline
$a_{ll}^0$ & $3.2187030369\dots$\\
\hline
$a_{\bar{l}L}^0$ & $5.8427632193\dots$\\
$a_{\bar{l}L}^1$ & $1.7461593992\dots$\\
\hline
$a_{lh}^0$ & $ 66.6327984972\dots$\\
$a_{lh}^1$ & $ 61.9658050968\dots$\\
\hline
$a_{\bar{h}L}^0$ & $45.6244273941\dots$\\
$a_{\bar{h}L}^1$ & $45.6244273770\dots$\\
$a_{\bar{h}L}^2$ & $5.4770385951\dots$\\
\hline
\end{tabular}
\end{minipage}
\begin{minipage}{0.45\textwidth}
\begin{tabular}{| c | l |}
\hline
Param.$/\langle\zeta\rangle$ & \\
\hline
$F_{l}^Z$ & $2.2927380937\dots$\\
$F_{h}^Z$ & $-1.6446346242\dots$\\
\hline
$b_{\bar{l}\bar{l}}^0$ & $18.4497693307\dots$\\
$b_{\bar{l}\bar{l}}^1$ & $12.8748121373\dots$\\
\hline
$b_{lL}^0$ & $46.9842992716\dots$\\
$b_{lL}^1$ & $42.1249702982\dots$\\
\hline
$b_{\bar{l}\bar{h}}^0$ & $ 695.4260997639\dots$\\
$b_{\bar{l}\bar{h}}^1$ & $ 960.8141050048\dots$\\
$b_{\bar{l}\bar{h}}^2$ & $ 276.4554994639\dots$\\
\hline
$b_{hL}^0$ & $531.5286291647\dots$\\
$b_{hL}^1$ & $851.3434042817\dots$\\
$b_{hL}^2$ & $325.5818137048\dots$\\
\hline
\end{tabular}
\end{minipage}
\caption{Parameters of the first FFs of the leading order operator $\sigma$ and subleading order operator $Z$.}
\label{t_leading_order_summary}
\end{table}

For the subleading order operator, the parametrization of the FFs is 
\begin{equation}
\begin{split}
F_{\bar{l}\bar{l}}^Z(\theta) &=\left(b_{\bar{l}\bar{l}}^0 +b_{\bar{l}\bar{l}}^1\cosh\theta\right) \frac{F_{\bar{l}\bar{l}}^{min}(\theta)}{D_{\bar{l}\bar{l}}(\theta)},\\
F_{lL}^Z(\theta) &=\left(b_{lL}^0+b_{lL}^1 \cosh\theta \right) \frac{F_{lL}^{min}(\theta)}{D_{lL}(\theta)},\\
F_{\bar{l}\bar{h}}^Z(\theta) &=\left(b_{\bar{l}\bar{h}}^0+b_{\bar{l}\bar{h}}^1 \cosh\theta+ b_{\bar{l}\bar{h}}^2 (\cosh\theta)^2\right) \frac{F_{\bar{l}\bar{h}}^{min}(\theta)}{D_{\bar{l}\bar{h}}(\theta)},\\
F_{hL}^Z(\theta) &=\left(b_{hL}^0+b_{hL}^1 \cosh\theta+ b_{hL}^2 (\cosh\theta)^2 \right) \frac{F_{hL}^{min}(\theta)}{D_{hL}(\theta)} 
\end{split}
\end{equation}
and the set of pole equations is the same as the one of $\sigma$, provided that we substitute $\sigma\to Z$ and charge-conjugate all particles, since $\sigma$ and $Z$ have complex conjugate $Z_3$ phases. It turns out that, in solving the set of linear equations, $F_{l}^Z$ and $F_{h}^Z$ remain unknown. Conjointly solving 
\begin{equation}
\label{eq_clusterBlL_subleading}
\begin{split}
\underset{\theta\to\infty}{\lim} F_{lL}^{Z}(\theta) &= \frac{w_{lL}}{\langle\zeta\rangle}F_{l}^{Z}F_{L}^{\zeta},\\
\underset{\theta\to\infty}{\lim} F_{hL}^Z(\theta) &= \frac{w_{hL}}{\langle\zeta\rangle}F_{h}^{Z}F_{L}^{\zeta},\\
\underset{\theta\to+\infty}{\lim} F_{l\bar{l}}^\zeta(\theta) &= \frac{w_{l\bar{l}}}{\langle \zeta\rangle}\left(F_{l}^Z\right)^2,
\end{split}
\end{equation}
allows us finding $F_{l}^Z$ and $F_{h}^Z$ without any phase ambiguity between them - $F_{l}^Z$ is fixed up to an overall sign, still. The parameters of the FFs of $Z$ are summarized in Table~\ref{t_leading_order_summary}.

%%%%%%%%%%%%%%%%%%%%%%

\section{Universal ratios of the renormalization group} \label{s6}
In the following we are going to consider some universal ratios of the renormalization group associated to the thermal deformation of  TPM, i.e. to the action eq.~\eqref{eq_thermally_deformed_TPM_action}. Consider an operator $\phi_j$ of the theory which has  at criticality the  anomalous dimension $\mathcal{X}_j = 2\Delta_j$ and  therefore renormalization group eigenvalue $y_j = 2-2\Delta_j$. The coupling $\lambda$ of the theory is a dimensional quantity related to the lightest excitation of the theory $m_l$ as in eq.~\eqref{eq_mass_lightestkink}. Denoting by $K_j$ the metric factor associated to the operator $\phi_j$, on general ground the scaling form of the VEV $\langle \phi_j \rangle$ can be written as (see \cite{fioravantiUniversalAmplitudeRatios2000} for a more detailed discussion) 
\begin{equation}
    \label{eq_scaling_expectation_values}
    \langle \phi_j \rangle = B_{j\epsilon} \lambda^{\mathcal{X}_j/y_\epsilon},
\end{equation}
with $B_{j\epsilon}:=K_j K_\epsilon^{\mathcal{X}_j/y_\epsilon}$. Generalized susceptibilities
\begin{equation}
\label{eq_scaling_generalized_susceptibilities}
    \hat{\Gamma}_{jk}^\epsilon = \Gamma_{jk}^\epsilon \lambda_\epsilon^{(2-y_j-y_k)/y_\epsilon},
\end{equation}
with $\Gamma_{jk}^i := K_j K_k K_\epsilon^{(2-y_j-y_k)/y_\epsilon}$,
are related to two-point connected off-critical correlators through the fluctuation-dissipation theorem
\begin{equation}
    \label{eq_dissipation_fluctuation}
    \hat{\Gamma}_{jk}^\epsilon = \int \dd^2 x \langle\phi_j(x)\phi_k(0)\rangle^{\text{C}}_\epsilon .
\end{equation}

Universal ratios are defined as ratios involving the correlation length $\xi$, the expectation values $\langle \phi_j \rangle$ and generalized susceptibilities $\hat{\Gamma}_{jk}^\epsilon$ such that the dependence on the metric factors $K_j$ cancels out \cite{fioravantiUniversalAmplitudeRatios2000}. These quantities  are associated to the class of universality and therefore are the same for different microscopic realizations of the model.  The first trivial universal ratio is the ratio between the correlation lengths in the disordered and ordered phase:
\begin{equation}
    \label{eq_univ_ratio_corr_func_above_below}
    \frac{\xi_+}{\xi_-} = \frac{m_L}{m_l} = 2\cos(\pi/4) = \sqrt{2} \approx 1.41421. 
\end{equation}
In order to find nontrivial universal ratios starting from the FFs computed in previous sections, eq.~\eqref{eq_dissipation_fluctuation} must be rewritten as to make the dependence on $\lambda$ manifest. Notice however that some susceptibilities are identically zero, because of $Z_3$-symmetry as, for instance, $\hat{\Gamma}^{\epsilon}_{\sigma\mu}$, $\hat{\Gamma}^{\epsilon}_{\sigma Z}$ and $\hat{\Gamma}^{\epsilon}_{\sigma\tilde{\sigma}}$. Charge conjugation also reduces the number of quantities to compute, e.g. $\hat{\Gamma}^{\epsilon}_{\tilde{\mu}\tilde{\mu}}=\hat{\Gamma}^{\epsilon}_{\mu\mu}$. In Table \ref{t_gen_susc_disorder} are presented the generalized susceptibilities involving $Z_3$-invariant operators ($\mu$, $\epsilon$ and $\zeta$), while Table \ref{t_gen_susc_order} is dedicated to magnetization ones ($\sigma$ and $Z$).
\begin{table}
\centering
\def\arraystretch{1.5}
\begin{tabular}{| c | l |l|l|l|l|l|}
\hline
State &
${ \hat{\Gamma}^{\epsilon}_{\mu\mu}\frac{ m_l^2}{\langle \mu\rangle^2}}$ & 
${ \hat{\Gamma}^{\epsilon}_{\epsilon\epsilon} \frac{ m_l^2}{\langle \epsilon\rangle^2}}$ &
${ \hat{\Gamma}^{\epsilon}_{\zeta\zeta}\frac{ m_l^2}{\langle \zeta\rangle^2}}$ &
${ \hat{\Gamma}^{\epsilon}_{\mu\epsilon} \frac{ m_l^2}{\langle \mu\rangle\langle \epsilon\rangle}}$ &
${ \hat{\Gamma}^{\epsilon}_{\mu\zeta}\frac{ m_l^2}{\langle \mu\rangle\langle \zeta\rangle}}$ &
${\hat{\Gamma}^{\epsilon}_{\epsilon\zeta}\frac{ m_l^2}{\langle \epsilon\rangle\langle \zeta\rangle}}$\\
\hline
\hline
$A_L$           & $0.12090$ & $0.90243$ & $5.05186$ 
                & $0.33031$ & $0.78153$ & $2.13517$\\
\hline
$A_{l\bar{l}}$  & $0.02860$ & $0.29656$ & $6.90261$
                & $0.09041$ & $0.35195$ & $1.19861$\\
\hline
$A_H$           & $0.00093$ & $0.01296$ & $0.18053$
                & $0.00347$ & $0.01296$ & $0.04837$\\
\hline
$A_{LL}$        & $0.00110$ & $0.02509$ & $0.53618$
                & $0.02329$ & $0.02329$ & $0.11530$\\
\hline
$A_{l\bar{h}}$  & $0.00237$ & $0.04371$ & $1.90493$
                & $0.01005$ & $0.05368$ & $0.24416$\\
\hline
\hline
Sum          &  $0.15390$ & $1.28076$ & $14.5761$
               &  $0.45754$ & $1.22341$ & $3.74160$\\
\hline
\end{tabular}
\caption{First contributions for the generalized susceptibilities of $Z_3$-invariant operators.}
\label{t_gen_susc_disorder}
\end{table}
\begin{table}
\centering
\def\arraystretch{1.5}
\begin{tabular}{| c | l |l|l|}
\hline
State &
${ \hat{\Gamma}^{\epsilon}_{\sigma\sigma} \frac{ m_l^2}{\langle \mu\rangle^2}}$ & 
${ \hat{\Gamma}^{\epsilon}_{\tilde{Z}\tilde{Z}} \frac{ m_l^2}{\langle \zeta\rangle^2}}$ &
${ \hat{\Gamma}^{\epsilon}_{\sigma\tilde{Z}} \frac{ m_l^2}{\langle \mu\rangle\langle \zeta\rangle}}$\\
\hline
\hline
$A_{\bar{l}}$  & $0.75482$ & $10.5133$ & $2.81702$\\
\hline
$A_{\bar{h}}$  & $0.01714$ & $1.44951$ & $0.15762$\\
\hline
$A_{ll}$       & $0.00838$ & $4.49913$ & $0.13430$\\
\hline
$A_{\bar{l}L}$ & $0.00919$ & $2.17994$ & $0.13840$\\
\hline
$A_{l h}$      & $0.00120$ & $1.33517$ & $0.02814$\\
\hline
$A_{\bar{h}L}$ & $0.00041$ & $0.33581$ & $0.01113$\\
\hline
\hline
Sum            & $0.79113$ & $20.3128$ & $3.28662$\\
\hline
\end{tabular}
\caption{First contributions for the generalized susceptibilities of magnetization operators.}
\label{t_gen_susc_order}
\end{table}

In general, FFs depend naturally on the VEV of some operator
\begin{equation}
    \label{eq_FF_redefinition}
    F^{\phi_j}_{a_1,\dots,a_n}(\theta_1,\dots,\theta_n) = f^{\phi_j}_{a_1,\dots,a_n}(\theta_1,\dots,\theta_n) \langle \phi_j' \rangle_i,
\end{equation}
where $\phi_j'$ is often coincident with $\phi_j$, like in the case of $\mu$, $\epsilon$ and $\zeta$, but not always, as it happens for $\sigma$ and $Z$. Manipulations of eq.~\eqref{eq_correlation_function_offcritical} yield
\begin{equation}
    \label{eq_spectral_Gamma}
    \Gamma_{jk}^\epsilon = 2\pi C_\epsilon^{-2} (B_{j\epsilon}')^* B_{k\epsilon}' \mathcal{F}_{jk}^\epsilon,
\end{equation}
where 
\begin{equation}
    \label{eq_spectral_series_adimensional}
    \mathcal{F}_{jk}^i = \int_0^\infty \dd\tau\ \tau \Biggl\lbrace \sum_{\lbrace n\rbrace} \int_{-\infty}^\infty \frac{\dd\theta_1 \dots \dd\theta_n}{n! (2\pi)^n} 
    \exp\left(-\tau \sum \beta_{a_i}\cosh\theta_i\right)  \left[ f^{\phi_j}_{a_1,\dots,a_n}(\theta_1,\dots,\theta_n)\right]^* f^{\phi_k}_{a_1,\dots,a_n}(\theta_1,\dots,\theta_n)
    \Biggr\rbrace.
\end{equation}
Here $\beta_{a_j}=m_{a_j}/m_l$. By approximating the two-point correlation function with the sum of the contributions associated to the one and two-particle states with lowest center-of-mass energy, we present hereafter some universal ratios of the thermal deformation of the TPM:
\begin{align}
    (R_c)^{\epsilon}_{\sigma \sigma} = \frac{\Gamma_{\epsilon \epsilon}^{\epsilon} \Gamma_{\sigma \sigma}^{\epsilon}}{B_{\mu\epsilon} B_{\mu\epsilon}}&\approx
    0.0153975135, \\
    (R_c)^{\epsilon}_{\sigma Z} = \frac{\Gamma_{\epsilon \epsilon}^{\epsilon} \Gamma_{\sigma Z}^{\epsilon}}{B_{\mu\epsilon}  B_{\tilde{\zeta}\epsilon}}&\approx
    0.0639662952, \\
    (R_c)^{\epsilon}_{\mu \mu} = \frac{\Gamma_{\epsilon \epsilon}^{\epsilon} \Gamma_{\mu \mu}^{\epsilon}}{B_{\mu\epsilon}  B_{\mu\epsilon}}&\approx
    0.0029953833, \\
    (R_c)^{\epsilon}_{\mu \epsilon} = \frac{\Gamma_{\epsilon \epsilon}^{\epsilon} \Gamma_{\mu \epsilon}^{\epsilon}}{B_{\mu\epsilon}  B_{\epsilon\epsilon} }&\approx
    0.0089049910, \\
    (R_c)^{\epsilon}_{\mu \zeta} = \frac{\Gamma_{\epsilon \epsilon}^{\epsilon} \Gamma_{\mu \zeta}^{\epsilon}}{B_{\mu\epsilon}  B_{\zeta\epsilon} }&\approx
    0.0238108180, \\
    (R_c)^{\epsilon}_{\epsilon \epsilon} = \frac{\Gamma_{\epsilon \epsilon}^{\epsilon} \Gamma_{\epsilon \epsilon}^{\epsilon}}{B_{\epsilon\epsilon}  B_{\epsilon\epsilon} }&\approx
    0.0249269560,  \\
    (R_c)^{\epsilon}_{\zeta \epsilon} = \frac{\Gamma_{\epsilon \epsilon}^{\epsilon} \Gamma_{\zeta \epsilon}^{\epsilon}}{B_{\zeta\epsilon} B_{\epsilon\epsilon}}&\approx
    0.0728215391, \\
    (R_c)^{\epsilon}_{Z Z} = \frac{\Gamma_{\epsilon \epsilon}^{\epsilon} \Gamma_{Z Z}^{\epsilon}}{B_{\tilde{\zeta}\epsilon} B_{\tilde{\zeta}\epsilon}}&\approx
    0.3953414813, \\
    (R_c)^{\epsilon}_{\zeta \zeta} = \frac{\Gamma_{\epsilon \epsilon}^{\epsilon} \Gamma_{\zeta \zeta}^{\epsilon}}{B_{\zeta\epsilon}  B_{\zeta\epsilon}  }&\approx
    0.2836900244.
    \comment{,\\
    (R_\chi)^\epsilon_\epsilon = (R_A)^\epsilon_\epsilon &= ((R_c)^{\epsilon}_{\epsilon \epsilon})^{1/2}\approx
    0.1578827286.}
\end{align}

It should be noted that the universal ratios involving the subleading order and disorder operators are subject to more uncertainty than those containing more relevant operators. Indeed for less relevant operators, the spectral series converges more slowly, as evidenced by the $\Delta$-theorem contributions.

%%%%%%%%%%%%%%%%%%%%%%%%%

\section{Comparison with a Monte~Carlo study}
\label{sec:comparison_with_a_Monte_Carlo_study}

In this section we present some results coming from the numerical simulations of the thermal deformation of the tricritical three-state Potts model on isotropic square lattices of spacing $a$. Hereafter we denote the sizes of the lattice in the two directions as $L_t=aN_t$ and $L_x=aN_x$; for all of the simulations described here, we used $L_t=L_x \equiv L$. Periodic boundary conditions are imposed in both directions; this choice, which preserves translational symmetry under translations by integer multiples of the lattice spacing along the $t$ and $x$ axes, minimizes the impact of artefacts due to the finiteness of the system. The reduced Hamiltonian of the lattice model is defined by eq.~\eqref{eq_diluted_potts_nienhuis}, which here we re-write in the following form,
\begin{align}
\label{reduced_Hamiltonian}
\frac{H}{\kB T} = -K \sum_{\langle i, j \rangle} \delta_{s_i,s_j}(1-\delta_{s_i,0}) - V \sum_{\langle i, j \rangle} \delta_{s_i,0}\delta_{s_j,0} -D \sum_i \delta_{s_i,0},
\end{align}
to account for the implementation used in our simulation code, where the spin and vacancy variables are simultaneously encoded into a single variable $s_i \in \left\{ 0, 1, 2, 3 \right\}$ defined on each lattice site $i$, with the value $0$ corresponding to a vacancy, while the other three values correspond to the site being occupied by a spin. As in eq.~\eqref{eq_diluted_potts_nienhuis}, the $\langle i, j \rangle$ notation denotes pairs of nearest-neighbor lattice sites. To simplify our numerical study, we restricted our attention to simulations for the $V=0$ case only -- that is, we neglected the possibility of a vacancy-vacancy coupling -- so that the dynamics of the model depends only on the two parameters $K$ (the coupling between nearest-neighbor spins) and $D$ (which acts as a chemical potential for the vacancies).

For our Monte~Carlo study, we created a dedicated C++ simulation algorithm, which generates the configurations of $s_i$ variables combining Swendsen-Wang cluster updates~\cite{Swendsen:1987ce} (for the spins) with local Metropolis updates~\cite{Metropolis:1953am} (for both the spins and the vacancies). One complete update consists of five Metropolis updates of the entire lattice alternated with five cluster updates of the whole system. While the non-local cluster updates dramatically reduce the critical slowing down close to critical points, this numerical problem is not completely solved, due to the persistence of non-negligible autocorrelations for the vacancies, whose evolution in Monte~Carlo time is affected only indirectly by the cluster updates.

We studied the model for a range of $K$ values in the proximity of the tricritical value $\Kcr=1.649903(5)$, holding $D$ fixed to its tricritical value $\Dcr=3.152152(2)$. These values for both $\Kcr$ and $\Dcr$ coincide with those reported in Ref.~\cite{Qian:2005dpm} where they were estimated by means of a different method, i.e. the transfer-matrix technique. We performed simulations at $81$ equally spaced values of $K$ in the $[\Kcr-0.04,\Kcr+0.04]$ interval; note that values of $K$ larger than $\Kcr$ correspond to thermal deformations driving the system into a broken-symmetry phase, in which the magnetization is expected to be non-zero in the thermodynamic limit. The square lattices that we simulated had sizes $(L/a)^2=64^2$, $128^2$, $256^2$, and $512^2$. For each value of $K$ and each lattice size, after at least $10^4$ complete updates for reaching Monte~Carlo thermalization, we collected $8\cdot10^5$ configurations for the lattices of sizes $(L/a)^2=64^2$ and $128^2$, $4\cdot10^5$ configurations for the lattices of size $(L/a)^2=256^2$, and a number of configurations between $387610$ and $751375$ for the lattices of size $(L/a)^2=512^2$. We used these ensembles of configurations to compute the expectation values of all observables described below, with the exception of the spin-spin correlation function, which was estimated from ensembles consisting of $4\cdot10^5$ configurations for each value of $K$ on the lattices of sizes $(L/a)^2=64^2$, $128^2$, and $(L/a)^2=256^2$, and a number of configurations between $201835$ and $386155$ for each $K$ value on the lattices of size $(L/a)^2=512^2$.

The observables that we evaluated in our Monte~Carlo simulations include: 
\begin{itemize}
\item The density of vacancies, which is the fraction of sites in the $s_i=0$ state
\begin{align}
\label{vacancy_density_definition}
\rho = \frac{1}{(L/a)^2}\sum_i \delta_{s_i,0}.
\end{align}
\item
The contribution to the energy (in units of $\kB T$) coming from spin-spin interactions, divided by the number of lattice sites, which is denoted as
\begin{align}
\label{energy_density_definition}
\epsilon = \frac{K}{(L/a)^2\kB T} \sum_{\langle i, j \rangle} \delta_{s_i,s_j}(1-\delta_{s_i,0}).
\end{align}
In the following (with a slight abuse of terminology), we will also refer to this quantity simply as `energy'; in addition, we considered the associated susceptibility, too.
\item 
The magnetization in each configuration, which  is defined as 
\begin{align}
\label{magnetization_definition}
\sigma = (1-\rho)\frac{qf_{\mbox{\tiny{max}}}-1}{q-1},
\end{align}
where $f_{\mbox{\tiny{max}}}$ denotes the fraction of spins in the majority state in the given configuration,
\begin{align}
\label{largest_spin_fraction_definition}
f_{\mbox{\tiny{max}}} = \mathrm{max}_{k\in\left\{1, 2, 3\right\}} \left( \frac{1}{(L/a)^2}\sum_i \delta_{s_i,k}\right).
\end{align}
\item
Finally, the two-point spin-spin correlation function, which is defined as
\begin{align}
\label{spispin_correlator_definition}
G_{s s}(r) = \frac{1}{2(L/a)^2}\sum_i \sum_{k=0}^1 \left(1-\delta_{s_i,0}\right)\delta_{s_i,s_{i+r\hat{u}_k}},
\end{align}
where $\hat{u}_0$ and $\hat{u}_1$ respectively denote the positively oriented unit vectors in the $t$ and $x$ directions of the lattice.
\end{itemize}
For all of these variables, we computed the integrated autocorrelation time (for a discussion about the numerical estimate of this quantity in Monte~Carlo simulations, see, e.g., Ref.~\cite{Wolff:2003sm}), and grouped the whole set of `measurements' of each observable obtained in every simulation into bins of length equal to the integrated autocorrelation time. We verified that, as expected, the average over each bin produced data with negligible residual autocorrelation.

\begin{figure}[!htb]
\begin{center}
\includegraphics*[width=0.5\textwidth]{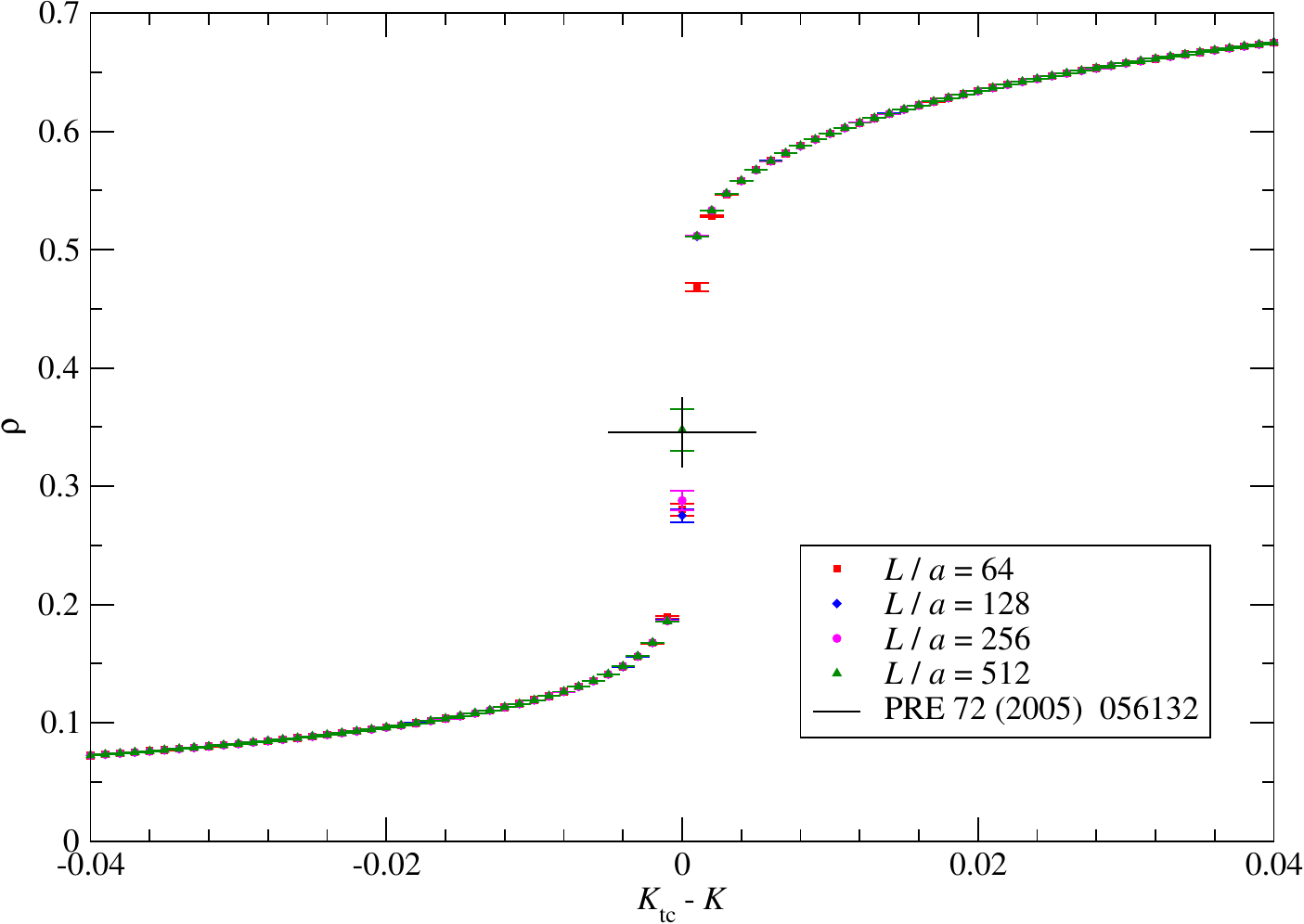}
\caption{Average density of vacancies, displayed as a function of $\Kcr-K$, obtained from our simulations on square lattices of different sizes (denoted by symbols of different shapes and colors). The plot also shows the average density of vacancies at $\Kcr$ (shown by the black cross) that was computed in Ref.~\cite{Qian:2005dpm} using a different numerical technique.\label{fig:average_vacancy_distribution}}
\end{center}
\end{figure}

Figure~\ref{fig:average_vacancy_distribution} shows how the distribution of vacancies extracted from our simulations varies as a function of $\Kcr-K$. The first observation is that the results from simulations on lattices of the different sizes that we considered exhibit a remarkable collapse of data, indicating that finite-volume corrections are under control. Furthermore, the data are consistent with an inflection point at $K=\Kcr$, and the value of $\rho$ at that point is in full agreement with the result (denoted by the black cross) obtained through the transfer-matrix technique in Ref.~\cite{Qian:2005dpm}.

%%%%%%%%%%%%%%%%%%%%%%%%%%%%%%%%%%%%%%%%%%%%%%

\begin{figure}[!htb]
\begin{center}
\includegraphics*[width=0.5\textwidth]{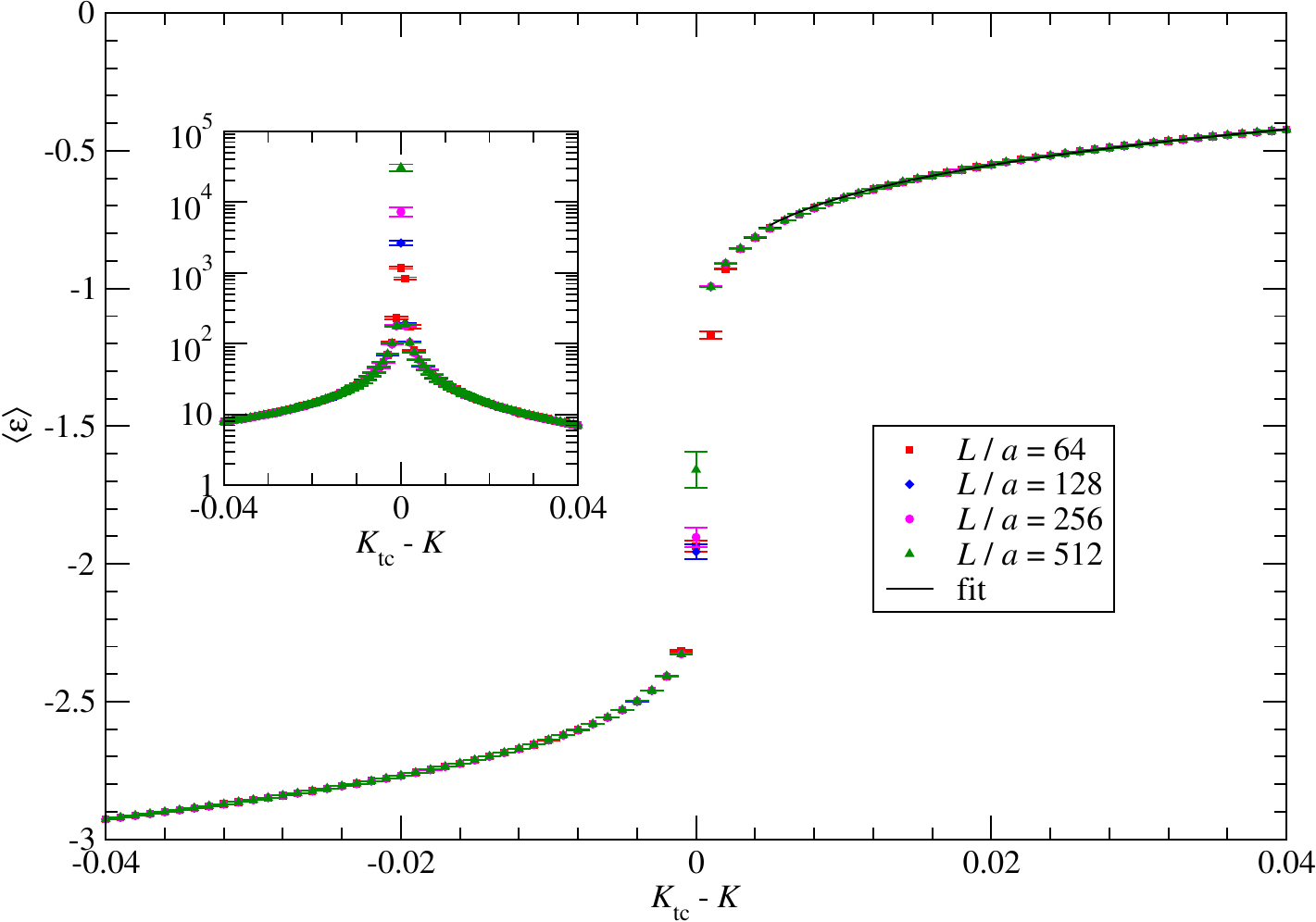}
\caption{Average contribution to the energy density from spin-spin interactions, plotted against $\Kcr-K$, from Monte~Carlo calculations on square lattices of linear size ranging from $L/a=64$ to $L/a=512$; the curve obtained by fitting the data for $0.005\le \Kcr -K \le 0.04$ to eq.~\eqref{energy_density_fit} is also shown. The inset plot shows the results for the susceptibility associated with $\epsilon$ (rescaled by the number of lattice sites), which are displayed with a logarithmic scale for the vertical axis.\label{fig:average_epsilon}}
\end{center}
\end{figure}

Next, we consider the average value of $\epsilon$, which is plotted in Fig.~\ref{fig:average_epsilon}, together with the results for the associated susceptibility (displayed in the inset plot, with a logarithmic scale on the vertical axis). As for the vacancy density, also for $\langle \epsilon \rangle$ we observe a clear collapse of the data obtained from simulations on lattices of different area, and an inflection point at a location consistent with $K=\Kcr$. For this observable, the Monte~Carlo results are expected to be modelled by a function of the form:
\begin{align}
\label{energy_density_fit}
\langle \epsilon \rangle = e_0\left(\Kcr-K\right)^{1/6}+e_1 .
\end{align}
Fitting the data in the range $0.005\le \Kcr -K \le 0.04$ to eq.~\eqref{energy_density_fit}, one obtains $e_0=2.036(4)$ and $e_1=-1.6124(23)$. The curve obtained from the fit is shown by the black line in Fig.~\ref{fig:average_epsilon}. The good quality of this fit, in which the exponent of $(\Kcr-K)$ is fixed to the value predicted analytically, confirms the validity of the conformal field theory description of the results obtained from Monte~Carlo simulations, at least for the range of $K$ values and for the lattice sizes under consideration in this study.

\begin{figure}[!htb]
\begin{center}
\includegraphics*[width=0.5\textwidth]{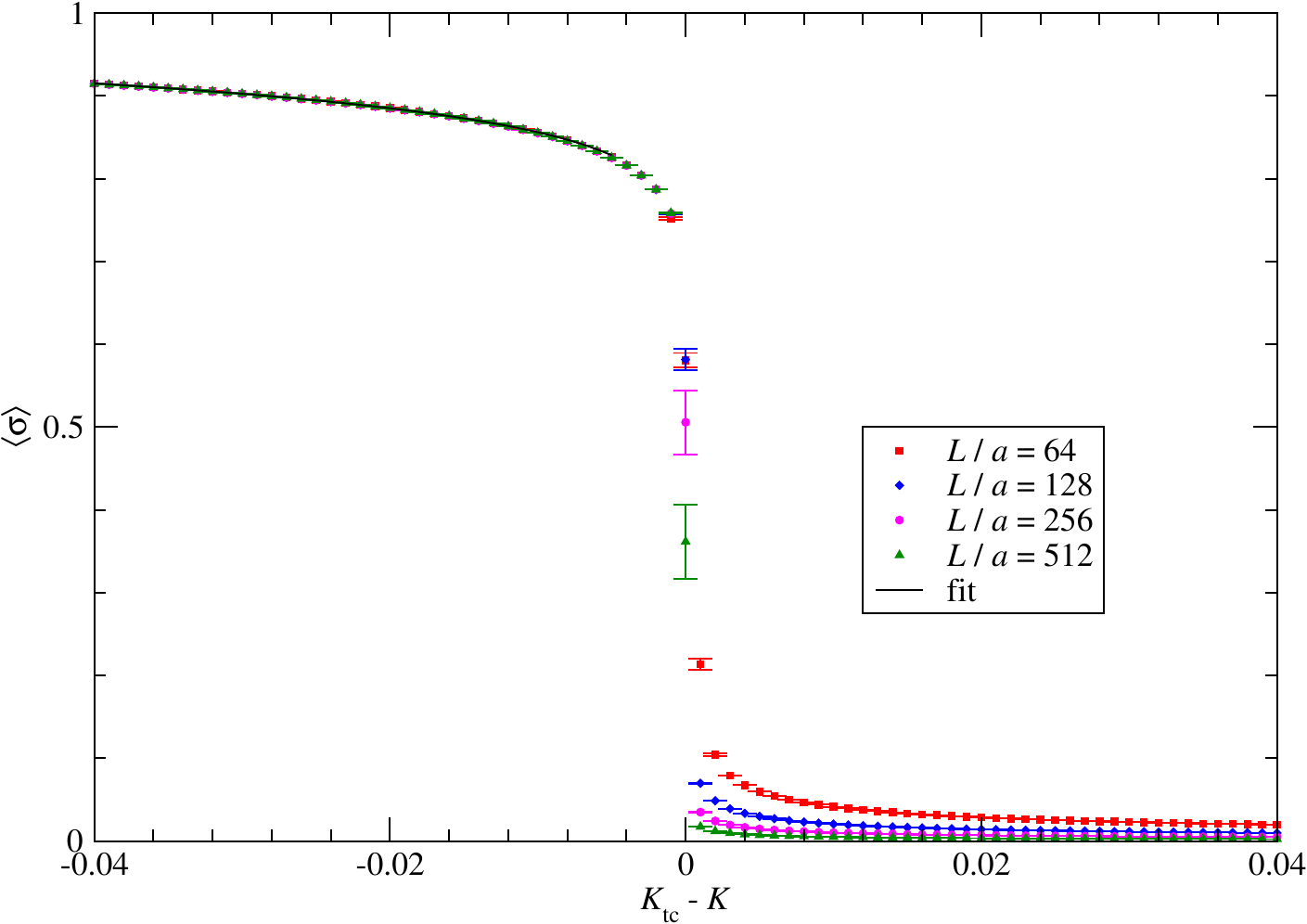}
\caption{Average magnetization, plotted against $\Kcr-K$, from our Monte~Carlo simulations on lattices with the same linear sizes as in Fig.~\ref{fig:average_vacancy_distribution} and in Fig.~\ref{fig:average_epsilon}. The curve obtained fitting our data in the broken-symmetry phase to eq.~\eqref{magnetization_fit} is also shown (black line). \label{fig:average_magnetization}}
\end{center}
\end{figure}

Our Monte~Carlo results for the average magnetization are displayed in Fig.~\ref{fig:average_magnetization}; note that, in the broken-symmetry phase, the results obtained from simulations on lattices of different area are consistent with each other. On the other hand, in the symmetric phase, the results obtained from smaller lattices are systematically larger than those from larger lattices: this is an artefact of the definition of the magnetization in eq.~\eqref{magnetization_definition}, which is based on the fraction of spins in the majority state in each configuration: by definition, for a system with a finite number of degrees of freedom, in the symmetric phase $f_{\mbox{\tiny{max}}}$ is always larger than $1/3$ -- except in the exceedingly rare configurations in which the total number of variables in the spin states $1$, $2$, and $3$ are exactly equal to each other. Nevertheless, our Monte~Carlo results show that the magnetization in the symmetric phase vanishes when it is extrapolated to the thermodynamic limit.

In the broken-symmetry phase, fitting the results for the magnetization in the range $-0.04 \le \Kcr-K \le -0.005$ to
\begin{align}
\label{magnetization_fit}
\langle \sigma \rangle = d_0\left|\Kcr-K\right|^{1/18}+d_1,
\end{align}
one obtains $d_0=0.9526(15)$ and $d_1=0.1185(13)$. The result of this fit is shown by the black line in Fig.~\ref{fig:average_magnetization}. As for $\langle \epsilon \rangle$, the fact that one obtains a good modelling of Monte~Carlo results using this simple functional form, with the exponent of $\left|\Kcr-K\right|$ fixed to the value that is predicted by conformal field theory, confirms the validity of this description of Monte~Carlo results. 

\begin{figure}[!htb]
\begin{center}
\includegraphics*[width=0.5\textwidth]{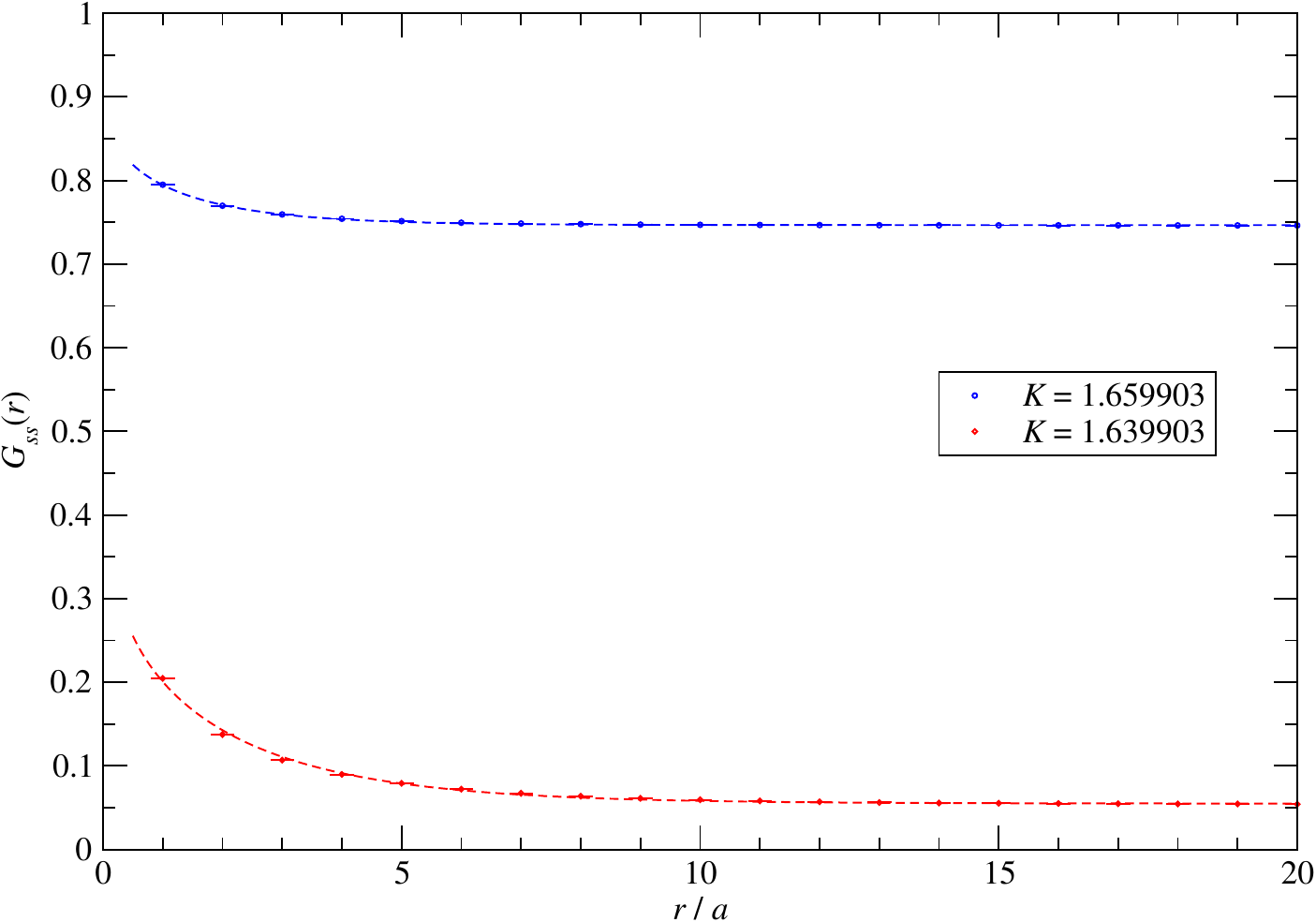}
\caption{Two-point spin correlation functions computed from Monte~Carlo simulations on a lattice of size $(L/a)^2=256^2$ at two different $K$ values, shown as a function of the spatial separation $r$ in units of the lattice spacing $a$. The dashed lines are the curves obtained from the fits to eq.~(\ref{average_spinspin_correlator_fit}).\label{fig:average_spinspin_correlator}}
\end{center}
\end{figure}

Next, we show examples of our results for the two-point spin-spin correlation function in Fig.~\ref{fig:average_spinspin_correlator}. In particular, it is interesting to study the universal ratio between the correlation lengths in the symmetric and in the symmetry-breaking phases. Sufficiently close to the tricritical point, it is expected that the ratio of correlation lengths obtained at values of $K$ that are symmetric with respect to $\Kcr$ should become consistent with the theoretical prediction $\xi_+/\xi_-=\sqrt{2}= 1.41421...$.

As an example to illustrate this type of study, the data displayed in Fig.~\ref{fig:average_spinspin_correlator} are obtained from two test ensembles of Monte~Carlo simulations on a lattice of size $(L/a)^2=256^2$ at $K=\Kcr\pm 0.01$, i.e., close to the tricritical point.\footnote{Each of these two additional ensembles consisted of $4\cdot10^5$ configurations, that were produced after $10^4$ complete updates for thermalization, like the other simulations on the lattices of size $(L/a)^2=256^2$. Note that, in order to (slightly) reduce the computing times, for these two test ensembles the normalization used in the Monte~Carlo calculation of the correlators explicitly dismissed an additive contribution to $G_{s s}(r)$; this additive contribution does not depend on $r$, and, hence, has no effect on the evaluation of the correlation lengths.} Expressing all distances in units of the lattice spacing, the correlators can be fitted to
\begin{align}
\label{average_spinspin_correlator_fit}
G_{s s}(r) = a_0\frac{\exp(-r/\xi)}{r^{4/21}}+a_2,
\end{align}
yielding $\xi=2.672(82)$ for $K=1.639903$ and $\xi=1.800(45)$ for $K=1.659903$, whose ratio
\begin{align}
\label{numerical_xiplus_over_ximinus_ratio}
\frac{\xi_{K=1.639903}}{\xi_{K=1.659903}} = 1.484(59)
\end{align}
is within less than $1.2$ standard deviations from the analytically predicted value.

\begin{figure}[!htb]
\begin{center}
\includegraphics*[width=0.5\textwidth]{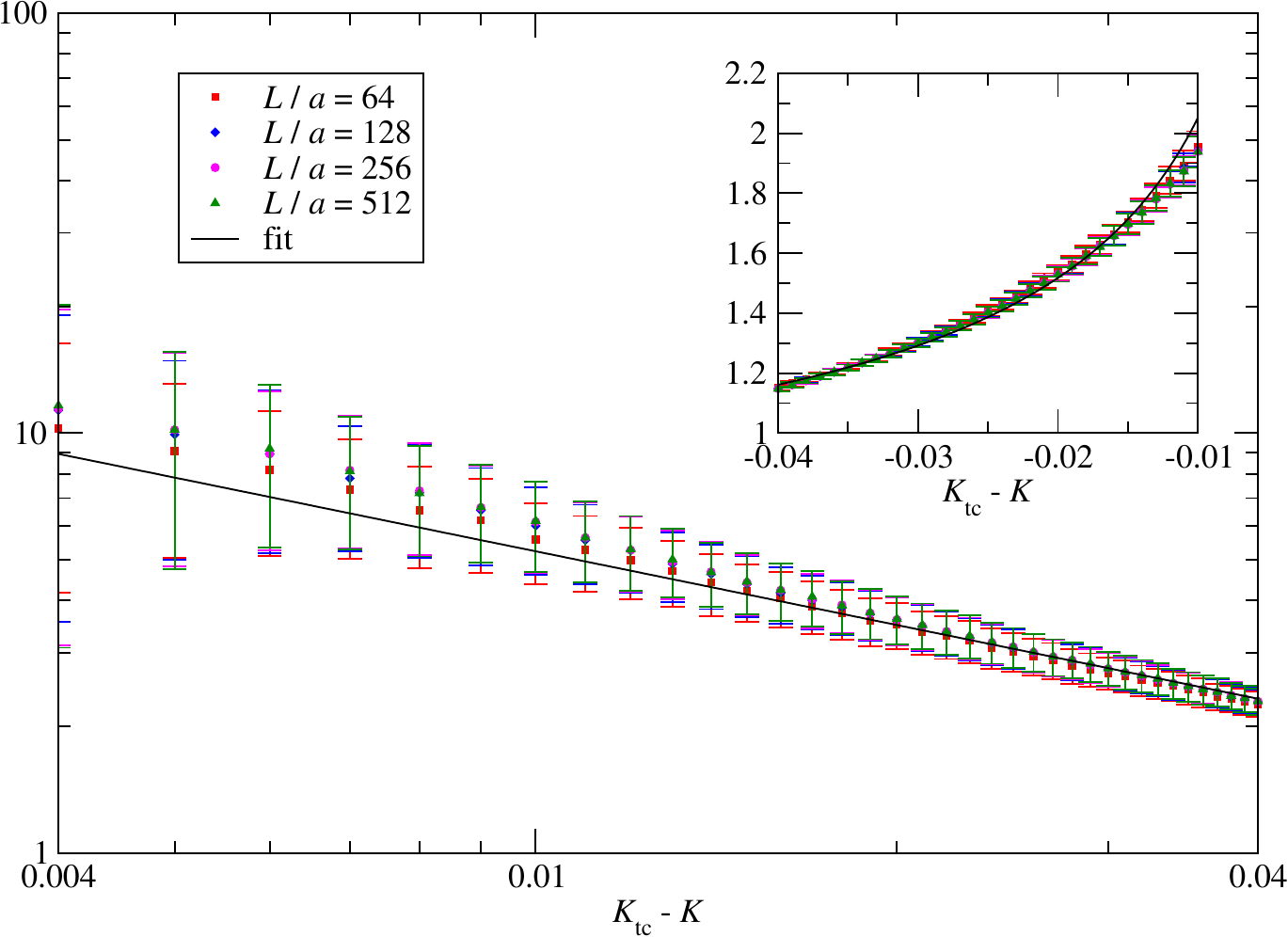}
\caption{Correlation lengths computed from Monte~Carlo simulations at different $K$ values and different lattice sizes, and their fit to the power-law behavior predicted from conformal field theory (black line), given by eq.~\eqref{correlation_length_fit}. The main plot refers to simulations at $K<\Kcr$, whereas the inset plot shows analogous results in the broken-symmetry phase. \label{fig:spinspin_correlation_lengths_and_fit}}
\end{center}
\end{figure}

It should be noted that the extraction of the correlation length from the Monte~Carlo results for the correlators is non-trivial, since it is based on a three-parameter, non-linear fit of cross-correlated data, and the fitted parameters are quite strongly correlated with each other, too. This, in turn, leads to a likely overestimate of the uncertainties affecting the fitted parameters, including, in particular, the correlation length. However, we followed a very conservative approach to the error budget, without attempting to reduce these uncertainties. Figure~\ref{fig:spinspin_correlation_lengths_and_fit} shows the correlation lengths computed from Monte~Carlo simulations on lattices of different sizes and at different $K$ values. In particular, the main plot (in which a logarithmic scale is used for both axes) displays the data in the symmetric phase and their fit to the power-law dependence on $|\Kcr-K|$ predicted from the theory:
\begin{align}
\label{correlation_length_fit}
\xi = a_0 |\Kcr-K|^{-7/12}.
\end{align}
The plot makes manifest that the simulation results precisely follow the power-law behavior predicted by conformal field theory. The only free parameter of the fit to eq.~\eqref{correlation_length_fit} is the overall amplitude, for which we obtain $a_0=0.356(1)$. Analogous results in the broken-symmetry phase are shown in the inset plot (where a linear scale is used for both axes).

%%%%%%%%%%%%%%%%%%%%%%%%%

\section{Discussion and conclusions}
In this work we have addressed the computation of the form factors of the relevant operators of the TPM once this model is perturbed away 
from its critical point by varying the temperature. The corresponding massive theory is integrable and has a distinguished fingerprint associated to the $E_6$ exceptional 
Lie algebra. We have determined the FFs associated to the leading and subleading order and disorder operators and we have checked their correct identification using the $\Delta$ sum rule. We have shown how to theoretically compute several universal ratios of the renormalization group associated to the class of universality of the model in any of its microscopic realizations. We also presented some results coming from a Monte~Carlo study of the lattice model, which provided clear evidence that, in the range of parameters that we considered, the results are in agreement with the expectations from CFT, and that, in addition, the universal ratio of the masses in the high- and low-temperature phases of the model is fully consistent with our theoretical prediction. 

\label{sec:discussion_and_conclusions}

%%%%%%%%%%%%%

\section*{Acknowledgements}
G.M. acknowledges the grants PNRR MUR Project PE0000023- NQSTI and PRO3 Quantum Pathfinder. The work of M.P. has been partially supported by the Spoke 1 ``FutureHPC \& BigData'' of the Italian Research Center on High-Performance Computing, Big Data and Quantum Computing (ICSC) funded by MUR (M4C2-19) -- Next Generation EU (NGEU), by the Italian PRIN ``Progetti di Ricerca di Rilevante Interesse Nazionale -- Bando 2022'', prot. 2022TJFCYB, and by the ``Simons Collaboration on Confinement and QCD Strings'' funded by the Simons Foundation. The simulations were run on CINECA computers. We acknowledge support from the SFT Scientific Initiative of the Italian Nuclear Physics Institute (INFN).

\pagebreak
\bibliographystyle{unsrtnat}
\bibliography{tpm}

\pagebreak
\appendix
\section{Analytic Properties of Form Factors} \label{a1}
In this Appendix, we will briefly touch upon some properties of FFs, complementing the discussion of Section \ref{s3}. Let us start with the connection between the matrix element in the r.h.s. of eq.~\eqref{eq_ff_translation} and eq.~\eqref{eq_ff_defi}, given by crossing
\begin{equation}
F^\mathcal{O}_{a_1,\dots,a_m, \bar{a}_{m+1},\dots, \bar{a}_n}(\theta_1,\dots,\theta_m, \theta_{m+1}-i\pi,\dots, \theta_{n}-i\pi) = F^\mathcal{O}_{a_1,\dots,a_n}(\theta_{ij}; \theta_{Ak};\theta_{BC}),  
\end{equation}
where $1\leq i < j\leq m$, $m+1\leq k \leq m < A \leq n$ and $m+1\leq B < C\leq n$.

In analogy to the $S$-matrix, we expect FFs to be analytic functions in the strip ${0<\Im\theta_{ij}<2\pi}$ and to have poles in correspondence of bound states. To understand the analytic properties satisfied by FFs, it is useful to define the `Faddeev-Zamolodchikov' (FZ) algebra, which follows from the definition of the scattering matrix. The FZ algebra is an associative algebra of vertex operators which satisfies the following commutation relations involving the $S$-matrix:
\begin{equation}
\label{eq_fz_algebra}
\begin{split}
A_{a_i}(\theta_i)A_{a_j}(\theta_j) &= S_{ij}(\theta)A_{a_j}(\theta_j)A_{a_i}(\theta_i),\\
A^\dagger_{a_i}(\theta_i)A^\dagger_{a_j}(\theta_j) &= S_{ij}(\theta)A^\dagger_{a_j}(\theta_j)A^\dagger_{a_i}(\theta_i),\\
A_{a_i}(\theta_i)A^\dagger_{a_j}(\theta_j) &= S_{ij}(\theta)A^\dagger_{a_j}(\theta_j)A_{a_i}(\theta_i)+2\pi \delta_{a_i a_j}\delta(\theta),
\end{split}
\end{equation}
with $\theta = \theta_i-\theta_j$. Here $A_{a_i}(\theta_i)$ is the destruction operator and $A^\dagger_{a_i}(\theta_i)$ its creation counterpart. Also $a_i$ is a shorthand for the set of quantum numbers that distinguishes a particle from the others in the spectrum.

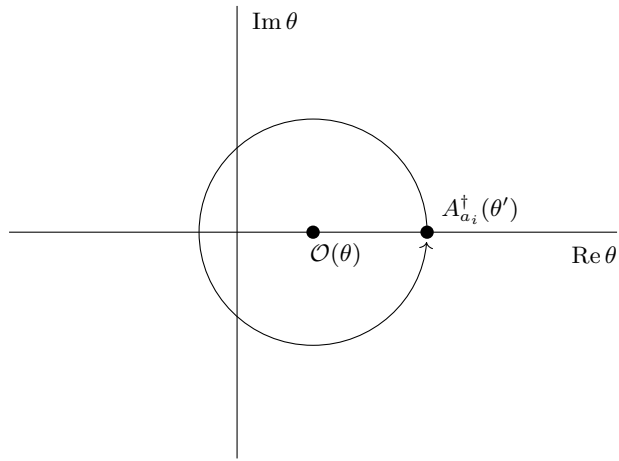
\begin{figure}
\centering
\makebox[\textwidth][c]{
\begin{tikzpicture}
  \node [ circle,fill=black,inner sep=0pt,minimum size=5pt]() at (0.0,0){};
  \node [ ]() at (0.3,-0.3){$\mathcal{O}(\theta)$};
  \coordinate [ ](xmax) at (4,0){};
  \coordinate [ ](xmin) at (-4,0){};
  \coordinate [ ](ymax) at (-1, +3){};
  \coordinate [ ](ymin) at (-1, -3){};
  \node [ ](re) at (3.7,-0.3){$\Re \theta$};
  \node [ ](im) at (-0.5, 2.8){$\Im \theta$};
  \draw[ ] (xmin.0) -- (xmax.0);
  \draw[ ] (ymin.0) -- (ymax.0);
  \draw[->] (1.5,0) arc (0:355:1.5);
  \node [ circle,fill=black,inner sep=0pt,minimum size=5pt]() at (1.5,0){};
  \node [ ]() at (2.2,0.3){$A^\dagger_{a_i}(\theta')$};
\end{tikzpicture}
}
\caption{The analytic continuation that defines non-local fields.}
\label{f_nonlocality}
\end{figure}

It is straightforward to rephrase the definition of FFs eq.~\eqref{eq_ff_defi} in the vertex operator formalism. Suppose that the vertex operator $A^\dagger_{a_i}$ is non-local with respect to the operator $\mathcal{O}$ \cite{zamolodchikovNonlocalParafermionCurrents1985}, which means that after a counterclockwise rotation of $A^\dagger_{a_i}(\theta')$ around $\mathcal{O}(\theta)$, the OPE ${A^\dagger_{a_i}(\theta')\mathcal{O}(\theta)}$ acquires a phase ${\exp{i2\pi \gamma_{a_i,\mathcal{O}}}}$. If $\gamma_{a_i,\mathcal{O}}$ is zero, then we have `mutual locality'. A graphical representation of this process is presented in Figure~\ref{f_nonlocality}.

Watson equations are direct consequences of the commutation relations eq.~\eqref{eq_fz_algebra} and the non-locality properties of the vertex operators with respect to $\mathcal{O}$:
\begin{equation}
\label{eq_first_watson}
F_{\dots,a_{m},a_{m+1},\dots}^{\mathcal{O}}(\dots,\theta_{m},\theta_{m+1},\dots) = S_{a_m,a_{m+1}}(\theta_m-\theta_{m+1})
F_{\dots,a_{m+1},a_{m},\dots}^{\mathcal{O}}(\dots,\theta_{m+1},\theta_{m},\dots),
\end{equation}
\begin{equation}
\label{eq_second_watson}
F_{a_1\dots a_n}^{\mathcal{O}}(\theta_1+2\pi i,\theta_{2},\dots,\theta_n) = e^{i2\pi \gamma_{a_1, \mathcal{O}}}F_{a_2\dots a_n,a_1}^{\mathcal{O}}(\theta_2, \dots,\theta_{n},\theta_1).
\end{equation}

From these equations it is clear that minimal FFs introduced in Section \ref{s3} are nothing more than building-block solutions for the two-particle Watson equation, if non-locality is not accounted for. For degenerate mass spectra with reflectionless, i.e. diagonal, $S$-matrix, the general structure of a two-particle amplitude is 
\begin{equation}
S_{ij}(\theta) = (-1)^{\delta_{i,\bar{j}}}\prod_{x\in\mathcal{A}_{ij}} s_x(\theta),
\end{equation}
where $s_x(\theta)$ is the building-block function eq.~\eqref{buildingblockk} 
and $\mathcal{A}_{ij}$ is the set of poles associated to the scattering amplitude - see e.g. Table \ref{t_smatrix_TPM}. Minimal form factors can be parametrized as
\begin{equation}
\label{eq_minimal_ff_generalform}
F_{ij}^{min}(\theta) = \left[-i\sinh\left(\frac{\theta}{2}\right)\right]^{\delta_{ij}}\prod_{x\in\mathcal{A}_{ij}}\left[h_x(\theta)\right]^{p_x},
\end{equation}
where $p_x$ is the multiplicity associated to $x\in\mathcal{A}_{ij}$, which coincides with the order of the pole in the corresponding $S$-matrix element. The building-block functions $h_x(\theta)$ satisfy
\begin{equation}
\label{eq_h_equation}
h_x(\theta) =-s_x(\theta)h_x(-\theta),
\quad 
h_x(\theta+2\pi i) =h_x(-\theta),
\end{equation}
and their minimal solution is
\begin{equation}
\label{eq_h_integral_representation}
h_x(\theta) = \exp{2\int_0^\infty \frac{\dd t}{t}\frac{\sinh((1-x)t)}{\sinh^2(t)}\sinh^2\left(\frac{\hat{\theta} t}{2i\pi}\right)},
\end{equation}
with $\hat{\theta} = i\pi-\theta$. It should be noted that $h_x(\theta)$ has poles on the imaginary axis of $\theta$, as it can be clearly see after performing an analytic continuation. There are two equivalent ways: the first is
\begin{equation}
\label{eq_h_product_gammas}
h_x(\theta) = \prod_{k=0}^{\infty} \frac{\left[\Gamma\left(k+\frac{1}{2}+\frac{x}{2}\right)\right]^2\Gamma\left(k+1-\frac{x}{2}-\frac{i\theta}{2\pi}\right)\Gamma\left(k+2-\frac{x}{2}+\frac{i\theta}{2\pi}\right)}{\left[\Gamma\left(k+\frac{3}{2}-\frac{x}{2}\right)\right]^2\Gamma\left(k+\frac{x}{2}-\frac{i\theta}{2\pi}\right)\Gamma\left(k+1+\frac{x}{2}+\frac{i\theta}{2\pi}\right)}, 
\end{equation}
while the second reads
\begin{equation}
\label{eq_h_product_logs}
h_x(\theta) = \prod_{k=0}^{\infty} \left[\frac{1+\left(\frac{\hat{\theta}/2\pi}{k+\frac{1}{2}+\frac{x}{2}}\right)^2}{1+\left(\frac{\hat{\theta}/2\pi}{k+\frac{3}{2}-\frac{x}{2}}\right)^2}\right]^{k+1}.
\end{equation}
It is possible to find analytic continuations with a finite number of factors, useful in numerical evaluations
\begin{equation}
\label{eq_h_analytic_continuation_gammas}
\begin{split}
h_x(\theta) &= \prod_{k=0}^{N-1} \frac{\left[\Gamma\left(k+\frac{1}{2}+\frac{x}{2}\right)\right]^2\Gamma\left(k+1-\frac{x}{2}-\frac{i\theta}{2\pi}\right)\Gamma\left(k+2-\frac{x}{2}+\frac{i\theta}{2\pi}\right)}{\left[\Gamma\left(k+\frac{3}{2}-\frac{x}{2}\right)\right]^2\Gamma\left(k+\frac{x}{2}-\frac{i\theta}{2\pi}\right)\Gamma\left(k+1+\frac{x}{2}+\frac{i\theta}{2\pi}\right)}\times\\
&\quad\times\exp{2\int_0^\infty \frac{\dd t}{t}\frac{\sinh((1-x)t)}{\sinh^2(t)}e^{-2Nt}\sinh^2\left(\frac{\hat{\theta} t}{2i\pi}\right)},
\end{split}
\end{equation}
\begin{equation}
\label{eq_h_analytic_continuation_logs}
h_x(\theta) = \prod_{k=0}^{N+1} \left[\frac{1+\left(\frac{\hat{\theta}/2\pi}{k+\frac{1}{2}+\frac{x}{2}}\right)^2}{1+\left(\frac{\hat{\theta}/2\pi}{k+\frac{3}{2}-\frac{x}{2}}\right)^2}\right]^{k+1} \exp{2\int_0^\infty \frac{\dd t}{t}\frac{\sinh((1-x)t)}{\sinh^2(t)}\left(1+N-Ne^{-2t}\right)
e^{-2Nt}\sinh^2\left(\frac{\hat{\theta} t}{2i\pi}\right)}.
\end{equation}

The last property of $h_x(\theta)$ we mention is its asymptotic behavior \cite{acerbiFormFactorsCorrelation1996}, which finds a  natural application in the asymptotic factorization of FFs: 
\begin{equation}
\label{eq_asymptotic_behaviour_h}
h_x(\theta)\underset{|\theta|\to \infty}{\sim}\exp{\int_0^\infty\frac{\dd t}{t} \left[\frac{\sinh((1-x)t)}{\sinh^2(t)}-\frac{1-x}{t}\right]}\exp{\frac{1-x}{2}(\abs{\theta}-i\pi)}.
\end{equation}

Concerning the parametrization of FFs eq.~\eqref{eq_parametrization_ff}, kinematic poles can be implemented by adding a term ${\cosh(\theta_{i\bar{i}}/2)}$ at the denominator of the FF, while for the poles coming from the $S$-matrix one can use: 
\begin{equation}
\label{eq_D_general}
D_{ij}(\theta) = \prod_{x\in\mathcal{A}_{ij}}\left[P_x(\theta)\right]^{n_x},
\end{equation}
with
\begin{equation}
\label{eq_P_functions}
P_x(\theta) = \frac{\cos(\pi x)-\cosh\theta}{2\cos^2(\pi x/2)},
\end{equation}
and 
\begin{equation}
n_x = \begin{cases}
m+1, & \text{if } p_x = 2m+1 \text{ and the pole is in the s-channel;}\\
m, & \text{if } p_x = 2m+1 \text{ and the pole is in the t-channel;}\\
m, & \text{if } p_x = 2m.
\end{cases}
\end{equation}

%%%%%%%%%%%%%%%%%%%%%%%

\end{document}